\documentclass [aps,12pt,aps,a4paper,prb]{revtex4}
\usepackage{graphicx}  
\usepackage{dcolumn}   
\usepackage{bm}        
\usepackage{amssymb}   

\begin{document}



\title{Phase Separation in Symmetric Mixtures of Oppositely Charged Rodlike Polyelectrolytes}
\author{{\bf Rajeev Kumar, Debra Audus and Glenn H. Fredrickson\footnote[1]{To whom any correspondence should be addressed, Email : ghf@mrl.ucsb.edu} } }

\affiliation{Materials Research Laboratory, University of California, Santa Barbara, CA-93106-5080}
\date{\today}

\begin{abstract}
Phase separation in salt-free symmetric mixtures of oppositely charged rodlike
polyelectrolytes is studied using quasi-analytical calculations.
Stability analyses for the isotropic-isotropic and the
isotropic-nematic phase transitions in the mixtures are carried out
and demonstrate that electrostatic interactions favor nematic
ordering. Coexistence curves for the symmetric mixtures are also
constructed and are used to examine the effects of linear charge
density and electrostatic interaction strength on rodlike
polyelectrolyte complexation. It is found that the counterions are
uniformly distributed in the coexisting phases for low electrostatic
interaction strengths dictated by the linear charge density of the
polyelectrolytes and Bjerrum's length. However, the counterions also
partition along with the rodlike polyelectrolytes with an increase
in the electrostatic interaction strength. It is shown that the
number density of the counterions is higher in the concentrated (or
``coacervate'') phase than in the dilute (or supernatant) phase. In
contrast to such rodlike mixtures, flexible polyelectrolyte mixtures
can undergo only isotropic-isotropic phase separation. A comparison
of the coexistence curves for weakly-charged rodlike mixtures with
those of analogous flexible polyelectrolyte mixtures reveals that
the electrostatic driving force for the isotropic-isotropic phase
separation is stronger in the flexible mixtures.
\end{abstract}

\maketitle

\section{Introduction}
For numerous biological
processes\cite{dubin_review,pe_complex_reviews,vlad_muthu,shklovskii_virus}
and emerging technologies\cite{waite,bazan_biosensor}, complexation
between oppositely charged
polyelectrolytes\cite{voorn_review,veis_complexation, kabanov_work1,kabanov_work2,dautzenberg_work1,pogodina_work,dautzenberg_work2} is the
underlying fundamental phenomenon. However, our understanding of how
the electrostatic attraction between opposite charges on the
polyelectrolytes is coupled with the conformational characteristics
of polyelectrolyte chains is very limited. Motivated by a plethora
of relevant biological processes and the impetus for developing
advanced technologies such as underwater adhesives\cite{waite} and
biosensors\cite{bazan_biosensor},  extensive
experimental\cite{dubin_review,pe_complex_reviews,waite,bazan_biosensor}
and theoretical\cite{zhaoyang_complex,
voorn_review,veis_complexation,voorn_overbeek,voorn_1,voorn_2,voorn_3,voorn_4,voorn_5,
borue_complexation,monica_complexation,monica_complexation_2,joanny_01,
jay_yuri_1,jay_yuri_2} work has been carried out to understand the
mechanisms of polyelectrolyte charge complexation.

A few insights into the origin of polyelectrolyte complexation have been 
provided by simulations. Langevin dynamics simulations\cite{zhaoyang_complex} of two
oppositely charged polyelectrolyte chains have revealed that the
entropy of counterion release is the predominant driving force for
complexation in highly charged polyelectrolytes. In contrast, the
simulations show that direct electrostatic attractions drive the
association of two weakly charged polyelectrolytes whose counterions
are not condensed onto the chains. Such simulations involving only a
single pair of polymers are of course relevant only to extremely
dilute solutions of polyelectrolytes and are only suggestive of
complexation phenomena at finite concentrations. 
At such higher concentrations, complexation between oppositely
charged polyelectrolytes has the character of a phase separation
process in which a supernatant phase that is extremely dilute in
polymer macroscopically separates from a polyelectrolyte rich phase
that is  either a fluid (``complex coacervate'') or a solid
(``precipitate''). In this paper we shall be exclusively concerned
with systems that produce fluid-like polyelectrolyte complexes
properly classified as complex
coacervates.\cite{dubin_review,pe_complex_reviews,voorn_review,veis_complexation,
voorn_overbeek,voorn_1,voorn_2,voorn_3,voorn_4,voorn_5}
Extending conventional Langevin, Monte-Carlo, or molecular dynamics
particle-based simulations to polyelectrolyte mixtures at finite
concentrations has been hindered by the need for extremely large
computational resources to address the long-range character of the
electrostatic interactions and the inherently slow kinetics of the
dense coacervate phase and phase separation process.

Development of a theory to study the phase separation in a mixture 
of oppositely charged polyelectrolytes 
is a daunting problem due to the non-trivial coupling between long range effects such as
electrostatics and chain connectivity, and short range excluded
volume interactions. The theoretical treatment is further
complicated by the inevitable need to include electrostatic
screening and correlation effects in the theory, analogous to the
classical Debye-H\"{u}ckel theory of electrolyte solutions, to capture
the electrostatic forces that drive phase separation. This is the
reason behind the inability of the self-consistent field theory
(SCFT), a type of mean-field theory, to capture even the signature
of a phase
separation\cite{borue_complexation,monica_complexation,monica_complexation_2,joanny_01,
jay_yuri_1,jay_yuri_2}. In other words, in a field-theoretic
description of a mixed polyelectrolyte system, one needs to go
beyond the mean-field or ``saddle-point'' approximation in order to
study phase separation phenomena. Typically, the random phase
approximation\cite{borue_complexation,monica_complexation,monica_complexation_2}
is used to compute the leading correction to the saddle-point
results. In the case of flexible
polyelectrolytes\cite{muthu_double}, it can be shown rigorously that
the random phase approximation is valid in the dual limit of high
monomer densities and low small ion densities. Physically, such a
situation is realized in concentrated solutions of weakly charged
polyelectrolytes. More sophisticated field-theoretic results beyond
the random phase approximation have also been obtained (including a
phase diagram) for a particular model of flexible polyelectrolyte
mixtures using the self-consistent one-loop (Hartree)
approximation\cite{joanny_01}. Most recently, direct numerical
simulations of a related field theory model (so-called
``field-theoretic simulations'') have been carried
out\cite{jay_yuri_1,jay_yuri_2} to capture the effects of
fluctuations and correlations to all orders, without invoking any
approximation. An important result that came out of these
simulations of flexible polyelectrolyte mixtures is that the phase
boundaries in the concentrated regime are accurately predicted by the
random phase approximation.

In this work, we consider solutions containing \emph{binary
mixtures} of oppositely charged rodlike polyelectrolytes and study
their phase behavior using the random phase approximation, which, as
noted above, has been previously shown to provide an accurate
description of complexation phenomena in flexible polyelectrolyte
mixtures. An important focus of our study is a comparison of the
phase behavior of rodlike and flexible polyelectrolyte mixtures. The
rodlike system is particularly relevant to a recently developed
biosensor technology\cite{bazan_biosensor} involving complexation of
cationic conjugated polyelectrolytes with anionic DNA.

A fundamental question that we have tried to answer is: what is the
role played by the flexibility of the polyelectrolytes in the phase
separation processes that lead to a complex coacervate? Unlike
flexible polyelectrolyte
mixtures\cite{borue_complexation,monica_complexation,monica_complexation_2,joanny_01,jay_yuri_1,jay_yuri_2},
local enhancements in concentration of rodlike systems can produce
orientationally ordered liquid crystalline phases (such as nematic,
smectic and cholesteric phases).\cite{degennes_prost} Thus, a
coacervate produced by complexing oppositely charged rodlike
polymers may be a liquid crystal rather than an isotropic fluid.  In
this paper, where we consider ``symmetric'' achiral rodlike polymers
with fore-aft symmetry, equal length, and equal but opposite charge,
we have explored the competition between isotropic and nematic
ordering in rodlike polyelectrolyte mixtures and elucidated how the
phase boundaries depend on different parameters of the system.

Before presenting our theory and numerical results on the \emph{binary mixtures}, it is worthwhile to briefly review 
the theoretical literature aimed at understanding the phase separation of \emph{single component}
flexible\cite{degennes_pincus,joanny_leibler,khokhlov_nyrkova,muthu_weak_polymer,chilun_muthu}
and
rodlike\cite{onsager_paper,odijk_pe_rods,chen_koch,carri_muthu99,potemkin_1,potemkin_2,potemkin_3,stell_rods}
polyelectrolyte solutions. These studies have been primarily focused on poor solvent conditions. In the case
of solutions containing a flexible polyelectrolyte, it has been
shown that the dependence of the phase boundaries on the
polyelectrolyte chain length is weak (referred to as the ``weak
polymer effect''\cite{muthu_weak_polymer}). Note that the condition
of Donnan equilibrium was invoked in these calculations so that
locally the system is electroneutral. This implies that a local
increase in the charge on the polyelectrolyte leads to an increase
in the local concentration of counterions. The localisation of the
counterions is entropically unfavorable and hence, the region of
two-phase coexistence shrinks with an increase in the charge on the
polyelectrolyte chains\cite{joanny_leibler,khokhlov_nyrkova,muthu_weak_polymer}.

Phase-separating mixtures of \emph{two oppositely charged
polyelectrolytes} are fundamentally different from these solutions
containing a single kind of polyelectrolyte in two ways. Firstly,
oppositely charged polyelectrolyte mixtures can phase separate even
under good solvent conditions for the two individual polymer
components. Secondly, it is clear in the binary mixture case that
counterions are free to partition very differently upon phase
separation. In particular, Donnan equilibrium can be satisfied by
bringing the oppositely charged macromolecules together in the
concentrated coacervate
phase,\cite{voorn_1,voorn_2,voorn_3,voorn_4,voorn_5} while freeing
the small ions to gain entropy by populating both dilute and
concentrated phase in charge compensating proportions. Clearly, it
is necessary to account for these extra degrees of freedom available
to the counterions in studying phase separation in mixtures of
oppositely charged polyelectrolytes.

In the case of \emph{single component} rodlike polyelectrolyte
solutions, it has been shown
recently\cite{potemkin_1,potemkin_2,potemkin_3} that electrostatic
interactions among like-charged rods favors orientational ordering
at concentrations lower than the overlap concentration. This
non-trivial result is an outcome of long-range, multi-rod
correlations present in these systems. A consequence is that, in
addition to the lyotropic isotropic-nematic transition observed in
neutral rodlike solutions, \emph{thermotropic} isotropic-nematic as
well as nematic-nematic phase transitions have been predicted for
charged rodlike polyelectrolytes. Note that these predictions are a
consequence of collective phenomena at higher concentrations that
are not manifest in earlier theories of dilute rodlike
polyelectrolytes, where the twisting
effect\cite{onsager_paper,parsegian} of pairwise electrostatic
interactions between rods was shown to destroy nematic
ordering\cite{onsager_paper,odijk_pe_rods,chen_koch,carri_muthu99}.

With this background on the phase separation in polyelectrolyte solutions, we 
present our theoretical model for the binary mixtures. 
This paper is organized as follows: the theory is presented in
Section \ref{sec:theory}, the results are presented in Section
\ref{sec:results}, and Section \ref{sec:conclusions} contains our
conclusions.

\section{Theory } \label{sec:theory}

We consider a \textit{symmetric} binary mixture of rodlike polyelectrolytes
bearing opposite charges on their backbone in the presence of the
counterions originating from the polyelectrolyte chains. By
\textit{symmetric} mixtures, we mean mixtures containing equal
number of polyelectrolyte chains, which are identical to each other
in every aspect other than the sign of the charge they are bearing.
For generality in the following, we maintain notation representing different polymeric
species such as $n_+$ and $n_-$ representing the number of polycationic and polyanionic 
chains, resepectively. For the theoretical treatment, we consider a binary mixture of rods 
with the same diameters ($= d$) for each species but lengths $L_k$ for $k=+,-$. 
Furthermore, $\sigma_k$ is taken to be the linear charge density for
polyelectrolytes of type $k=+,-$ so that there are $n_{ck} = \sigma_k L_k$
counterions released by each polyelectrolyte of type $k$. As the
charge on the polymers is spread uniformly along their length, this
description is often referred to as the ``smeared charge'' model.
We specialize to the symmetric case only as needed after plugging $n_+ = n_- = n, L_+ = L_- =  L$ and $\sigma_+ = \sigma_- =  \sigma$ at the appropriate place. Also, we work in the canonical ensemble with the volume of the system denoted by $\Omega$.

The solvent is treated implicitly and as a uniform dielectric
background for the purpose of computing electrostatic interactions.
The small ions are taken to be point like and we ignore short range
structure, correlations, and polarization effects in the solvent.

\subsection{Qualitative Picture and Scaling Analysis} \label{sec:scaling_section}
Before presenting quantitative details of our theory for the
symmetric rodlike polyelectrolyte mixtures, a
qualitative picture can be drawn by using three important results
available from the literature. First, the macrophase separation in
the isotropic phase (by analogy with symmetric mixtures of flexible
polylectrolytes) takes place at very low monomer
densities\cite{borue_complexation,monica_complexation,monica_complexation_2,joanny_01,jay_yuri_1,jay_yuri_2},
when the electrostatic attractions between oppositely charged
polyelectrolytes are strong enough to drive the phase separation.
Second, the isotropic-nematic phase transition observed in neutral
rodlike molecules and treated by Onsager and many
others\cite{edwardsbook,degennes_prost,onsager_paper,flory_semiflex,flory_rods,
khokhlov_semiflex,khokhlov_semenov} takes place at relatively high
monomer densities above the overlap concentration and is driven
primarily by entropic effects associated with the excluded volume of
rodlike objects. Thirdly, we recall that addition of charges to
rodlike polymers can induce thermotropic (temperature dependent)
isotropic-nematic phase transitions, as opposed to the lyotropic
(concentration dependent) transitions familiar in neutral rodlike
polymer systems. The origin of this thermotropic behavior lies in
the long-ranged electrostatic interactions, which depend on the
charge on the polyelectrolytes and the temperature dependent Bjerrum's
length $l_B =  e^2/(4\pi \epsilon_0\epsilon_r k_B T)$, which is the
distance at which the Coulomb interaction energy between two
elementary charges is of the order of thermal energy ($=k_BT, \;
k_B$ being the Boltzmann's constant and $T$ is the temperature).
Here, $\epsilon_0$ and $\epsilon_r$ are the permittivities of vacuum
and the medium, respectively.

\begin{figure}[ht!]
\vspace*{-0.95cm}
\begin{center}
\includegraphics[width=3in,height=3in]{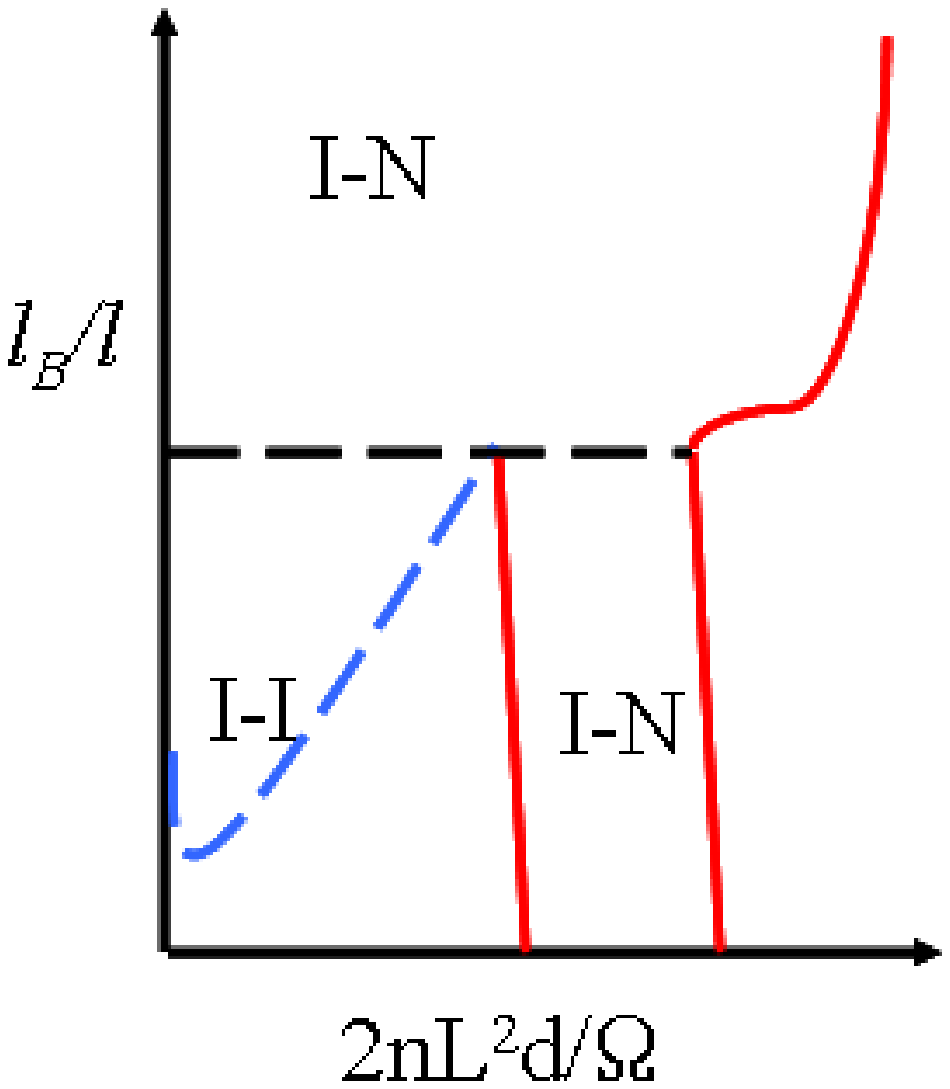}
\newline
\end{center}
\caption{A qualitative sketch of the phase diagram in a symmetric
mixture of oppositely charged rodlike polyelectrolytes. The symbols
``I'' and ``N'' denote regions of two-phase coexistence involving
isotropic and nematic phases, respectively. $2n/\Omega, \; L$ and
$d$ are the number density of positive and negative
polyelectrolytes, the length, and the diameter of the charged rods,
respectively. The y-axis is the ratio of the Bjerrum length $l_B$ to
the monomer length $l$ and is a measure of electrostatic strength
that is inversely proportional to temperature.}
\label{fig:phase_cartoon}
\end{figure}

By integrating these three competing effects, a \emph{qualitative}
phase diagram of a symmetric rodlike polyelectrolyte mixture is
sketched in Fig.~\ref{fig:phase_cartoon} in the coordinates of $l_B
/l$ versus $2 n L^2 d/\Omega$, where $l$ is a reference monomer
length, $L$ is the overall rod length, $d$ is the rod diameter, $2
n$ is the total number of cationic and anionic rods, and $\Omega$ is
the system volume. Broadly, we see two trends in the figure. At high
temperatures, i.e. low values of $l_B /l$, non-interacting
isotropic-isotropic  and isotropic-nematic coexistence regions
appear at low and high rod concentrations, respectively. As the
temperature is lowered ($l_B /l$ raised), these features collide and
ultimately at low temperature there is a broad region of two phase
coexistence between a dilute isotropic supernatant phase and a
semidilute or concentrated nematic ``coacervate'' phase. At
intermediate values of $l_B /l$, more complex phase behavior is
evident, including the possibility of three-phase coexistence
between two isotropic phases and a nematic phase.

With regard to the isotropic-isotropic coexistence region in
Fig.~\ref{fig:phase_cartoon}, which is also present in flexible
polyelectrolyte mixtures, a scaling analysis is helpful to
understand the variables that control its extent and shape and
differentiate the cases of rigid and flexible systems. For this
purpose, we begin by considering a symmetric mixture containing an equal
number $n$ of oppositely charged \emph{flexible} polyelectrolytes in
the absence of any counterions. Physically, such a situation is realized in a solution
containing polyacids and polybases. Apart from electrostatic
interactions, excluded volume effects can be simply accounted for
(in an implicit solvent model) by an excluded volume parameter $w$.
This particular system was recently studied using the
field-theoretic simulation technique\cite{jay_yuri_1,jay_yuri_2}.
The osmotic pressure of such a system in the dilute
regime\cite{degennes_pincus,rubinstein_review}, i.e. $\rho =
2nN/\Omega \ll \rho_f^\star \sim 1/(4\pi l_B \sigma^2 N^2)$, with
$\rho, \;\rho_f^\star$ being the monomer density and the overlap
concentration, respectively, is given by
\begin{eqnarray}
\frac{P_{f}^0}{k_BT} &=& \frac{\rho}{N} + \frac{w}{2} \rho^2 - \frac{\kappa_d^3}{24 \pi}, \label{eq:pressure_flex_dil}
\end{eqnarray}
where $\kappa_d^2 = 4\pi l_B \sigma^2 N \rho$ is the Debye screening
length in the asymptotic dilute regime where the polyelectrolytes
behave as multivalent macroions. We note that although the polymers
are assumed to be flexible, the expression that we use for
$\rho_f^\star$ assumes that they adopt extended rodlike
conformations at infinite dilution. In the above expression,
$\sigma$ is the charge per monomer (in units of the fundamental
charge $e$), which is the same for each kind of polyelectrolyte in
the assumed symmetric mixture and $N$ is the total number of
monomers.

In Eq. ~\ref{eq:pressure_flex_dil}, the first and second terms
correspond to the translational entropy of the polyelectrolyte
chains and the effect of excluded volume interactions on the
pressure, respectively. The third term containing $\kappa_d$, a
Debye-H\"{u}ckel contribution, originates from attractive
electrostatic correlations among dilute, non-overlapping
polyelectrolyte chains. In contrast, at higher polymer
concentrations where chains overlap the internal structure of the
polyelectrolytes plays a significant role. The dominant
electrostatic contribution to the osmotic pressure in the high
density, overlapping regime ($\rho \gg \rho_f^\star $) can be
represented by the scaling expression
\begin{eqnarray}
\frac{P_{f}^{e}}{k_BT} &=& - \frac{\kappa_d^3}{24 \pi} \:
\mathcal{F} \left(\rho / \rho_f^\star \right),
\end{eqnarray}
where $\mathcal{F} (x)$ is a dimensionless scaling function that we
assume to be a power law $\sim x^m$ for large argument. The exponent
$m$ is determined by the requirement that in the concentrated
regime, the local properties are not different for a solution having
multiple chains containing $N$ monomers each or a single chain that
fills space with the same overall segment density $\rho$
($N\rightarrow \infty$). In other words, the electrostatic
contribution to the osmotic pressure must be independent of $N$ at
fixed $\rho$, i.e. $P_f^e \sim N^0$. This requirement, along with
the fact that the overlap concentration scales as $\rho_f^\star \sim
1/(4\pi l_B \sigma^2 N^2)$, leads to the well-known result\cite{muthu_double,kaji_kanaya}
\begin{eqnarray}
\frac{P_{f}^{e}}{k_B T} &\sim& - \kappa_p^{3/2},
\label{eq:pressure_flex}
\end{eqnarray}
where $\kappa_p^2 = 4\pi l_B \sigma^2 \rho$. In the concentrated
regime, the osmotic pressure of flexible, symmetric polyelectrolyte
mixtures is thus dictated by a balance between excluded volume and
this modified electrostatic correlation energy, i.e.
\begin{equation}
\frac{P_f^c}{k_B T} \sim \frac{w}{2} \rho^2 - \kappa_p^{3/2}
\end{equation}


We can now retrace these scaling arguments for the case of symmetric
mixtures of \emph{rodlike} polylectrolytes. The osmotic pressure in
the dilute concentration regime ($\rho \ll \rho_r^\star \sim 2nL^2
d/\Omega$) is unchanged except that we replace the excluded volume
coefficient $w$ with the Onsager expression $\pi d l^2 /2$, where $l
\equiv L/N$ is a monomer segment length,
\begin{eqnarray}
\frac{P_{r}^0}{k_BT} &=& \frac{\rho}{N} + \frac{\pi d l^2}{4} \rho^2 - \frac{\kappa_d^3}{24 \pi}. \label{eq:pressure_rods_qual}
\end{eqnarray}
In the concentrated regime, the electrostatic free energy of a
system of rods can be approximated by the sum of the self-energies
of individual rods experiencing the electrostatic potential induced
by the other rods. For the purpose of estimation, these are placed
on a periodic lattice. It is well-known that the self-energy of a
charged cylindrical rod of length $L$ and linear charge density
$\sigma$ is divergent\cite{odijk_self,safran_pincus}, and is given
by
\begin{eqnarray}
\frac{U_{cyl}}{L k_BT} &\sim& l_B \sigma^2 \ln \frac{R}{\delta}, \label{eq:cyl_energy}
\end{eqnarray}
where $\delta$ is a cut-off introduced to regularize the self-energy
and $R$ is the radius of a unit cell (Wigner-Seitz) applied to each
cylinder. Using the fact that $R \sim 1/\sqrt{\rho}$ in the
concentrated regime and writing the electrostatic free energy of the
solution containing $n$ rods as $F^e  = n U_{cyl}$, the
electrostatic contribution to the pressure is given by
\begin{eqnarray}
\frac{P_{r}^e}{k_BT} &\sim& - \kappa_p^2.
\label{eq:pressure_rods_qual2}
\end{eqnarray}

Comparing Eqs. ~\ref{eq:pressure_flex} and
~\ref{eq:pressure_rods_qual2}, it is clear that the qualitative
difference between flexible and rodlike polyelectrolyte mixtures is
in the functional form for the attractive electrostatic contribution
to the osmotic pressure. In the case of rodlike polyelectrolytes,
this term scales like $\kappa_p^2 \sim \rho$ in contrast to
$\kappa_p^{3/2} \sim \rho^{3/4}$ for flexible coils. Thus, at the
same set of electrostatic parameters and monomer densities, and for
weakly charged polyelectrolytes where $\kappa_p \ll 1$,
electrostatic correlation effects are weaker for rodlike
polyelectrolytes in comparison with  flexible polyelectrolyte
mixtures. Since the electrostatic contribution to the pressure is
responsible for polyelectrolyte complexation, it follows that
rodlike polyelectrolyte mixtures are less prone to complex
coacervation than analogous flexible polyelectrolyte mixtures. The
regions of two-phase coexistence sketched in
Fig.~\ref{fig:phase_cartoon} should thus be narrower in rigid rod
systems. This general observation will be born out by detailed
numerical calculations in Section \ref{sec:results} of the spinodals
and binodals of each type of system based on the free energy expression 
presented below.

\subsection{Free Energy}
An expression for the Helmholtz free energy of the assumed mixture
of rodlike polyelectrolytes, counterions, and implicit solvent is
derived in Appendix A using Onsager's treatment for the neutral
interactions and the random-phase approximation for the
electrostatic interactions. Explicitly, the free energy is given by
\def\nh{\hat{\mathbf{n}}}
\begin{eqnarray}
      F &=& F_{en} + F_w  + F_e \label{eq:free_gen}
\end{eqnarray}
where
\begin{eqnarray}
   \frac{F_{en}}{k_BT} & = & \sum_{j=+,-}n_j \int d\mathbf{u}_{j} \: p_j(\mathbf{u}_{j},\nh) \ln \left[4\pi p_j(\mathbf{u}_{j},\nh)\right] + \sum_{\gamma=+,-,c+,c-} n_{\gamma} \left[\ln \frac{n_{\gamma}}{\Omega} - 1\right],\\
       \frac{F_w}{k_BT} &=&  \frac{1}{2\Omega} \sum_{j,k=+,-}^\prime n_j n_k \int d\mathbf{u} \int d\mathbf{u}' \: p_j(\mathbf{u},\nh)
\left [2L_j L_k d |\mathbf{u}\times\mathbf{u}'| \right]p_k(\mathbf{u}',\nh), \\
\frac{F_e}{k_BT} &=&   \frac{\Omega}{2}\int \frac{d^{3}\mathbf{q}}{(2\pi)^3} \left\{\ln \left[1 + \frac{\kappa^2
+ \sum_{k=+,-}\kappa_k^2 t_k(q L_k)}{q^2}\right] -  \frac{\kappa^2
+ \sum_{k=+,-}\kappa_k^2 t_k(qL_k)}{q^2}\right\}\label{eq:free_rods}
\end{eqnarray}
and
 \begin{eqnarray}
    t_k(qL_k)  &=& \int d\mathbf{u} \: p_{k}(\mathbf{u},\nh)\left[\frac{\sin\left[\left(\mathbf{q}\cdot\mathbf{u}\right)L_k/2\right]}{\left[\left(\mathbf{q}\cdot\mathbf{u}\right)L_k/2\right]}\right]^2, \quad \mbox{for} \quad k = +,-.
\end{eqnarray}

Here, $\kappa$ is the inverse of the Debye screening length for
small ions given by $\kappa^2 = 4\pi l_B \sum_{\gamma=c+,c-}
Z_{\gamma}^2 n_{\gamma}/\Omega$ so that $Z_\gamma$ is the valency (with sign) 
of the charged species of type $\gamma$. We define a similar object for the
two polymer species: $\kappa_k^2 = 4\pi l_B Z_{k}^2 \sigma_{k}^2
L_k^2 n_k/\Omega, Z_k$ being the valency of the charged monomers of type $k$. The functions $p_{k}(\mathbf{u},\nh)$ represent
the probability distribution function for finding a rod of type $k$
oriented along the unit vector $\mathbf{u}$ when the director is
chosen to be the unit vector $\nh$. Each distribution function
satisfies the normalization condition $\int d \mathbf{u} \:
p_{k}(\mathbf{u},\nh) = 1 $. Finally, the primed superscript in the
expression for $F_w$ indicates that the $j=k$ terms are omitted from
the double sum.


So far, the theory is quite general and the subsequent analysis
depends on the functional form for the orientational probability
distribution function for the rods. In this work, we focus on
possibilities for phase separation that involve the isotropic and
nematic phases only. For these two phases, $p_{k}(\mathbf{u},\nh)
\equiv p_{k}(\mathbf{u} \cdot \nh) $ and the free energy for the
each phase can be evaluated as described below.

\subsection{Isotropic Phase}

We begin by considering the isotropic phase, in which case the
orientation distribution function for the rods is independent of the
angle between the director $\nh$ and the unit vector along the axes
of the rods ($\mathbf{u}$). In this case, $p_{k}(\mathbf{u}\cdot
\nh) = 1/4\pi$ (obtained from the normalization condition) for
$k=+,-$ so that the free energy in Eq. ~\ref{eq:free_gen} can be
written as
\begin{eqnarray}
      F^{iso} &=& F_{en}^{iso} + F_{w}^{iso} + F_{e}^{iso} \label{eq:free_iso}
\end{eqnarray}
where
\begin{eqnarray}
   \frac{F_{en}^{iso}}{k_BT} & =& \sum_{\gamma = +,-,c+,c-} n_{\gamma} \left[\ln \frac{n_{\gamma}}{\Omega} - 1\right],\\
  \frac{F_w^{iso}}{k_BT} &=& \frac{\pi d}{4\Omega}(n_+ L_+ + n_- L_-)^2,  \\
 \frac{F_e^{iso} }{k_BT} &=& \frac{\Omega}{2}\int \frac{d^{3}\mathbf{q}}{(2\pi)^3} \left\{\ln \left[1 + \frac{\kappa^2 + \sum_{k=+,-} \kappa_k^2 t^{iso}(qL_k)}{q^2}\right] -   \frac{\kappa^2 + \sum_{k=+,-} \kappa_k^2 t^{iso}(qL_k)}{q^2}\right\} \label{eq:feiso}
\end{eqnarray}
and
 \begin{eqnarray}
 t^{iso}(qL_k)  &=& \frac{2Si(qL_k)}{q L_k} - \left(\frac{2\sin \left(qL_k/2\right)}{qL_k}\right)^2, \label{eq:tkq_iso}
\end{eqnarray}
 where $Si(x) = \int_{0}^{x}dt \frac{\sin t}{t}$ is the sine integral. 
 
 In the following, we rigorously show 
 that the scaling argument presented in section ~\ref{sec:scaling_section} predicting a logarithmic correction to the  free energy of isotropic phase in the 
 mixtures of rods is indeed correct. For the analysis, we use the following asymptotic expressions for $t^{iso} (q L_k )$ :
 \begin{equation}
 t^{iso}(qL_k)  = \left \{ \begin{array}{ll}
1, & qL_k \rightarrow 0 \\
 \pi/qL_k, &  qL_k \rightarrow \infty.
\end{array}
   \right . \label{eq:tkq_lim}
\end{equation}
Also, we consider two limiting cases of short and long rods. For these cases, the electrostatic contribution to the free energy depends on how these limits are being approached. These limits can be approached by either fixing the linear charge density or the net charge per rod during the variation of length $L_k$. In the former case, $\kappa_k$ increases (decreases) with an increase (decrease) in $L_k$ and diverges (vanishes) strictly in the limit of $L_k\rightarrow \infty$ ($L_k \rightarrow 0$). On the other hand, for the latter approach, $\kappa_k$ is held fixed, while approaching the limits of $L_k \rightarrow \infty$ or $L_k \rightarrow 0$. Consider the latter case so that $\kappa_k$ is well-defined in approaching either limits of $L_k$.

Defining a Debye-like parameter proportional to the total ionic
strength, $\kappa_{eff}^2 = \kappa^2 + \sum_{k=+,-} \kappa_k^2 $ and approaching the limit of short rods, $L_k \rightarrow 0$ while keeping the charge per rod fixed, the free energy becomes
 \begin{eqnarray}
 \frac{F^{iso}\left\{L_k \rightarrow 0\right\}}{k_BT}  &=& \sum_{\gamma = +,-,c+,c-} n_\gamma \left[\ln \frac{n_\gamma}{\Omega} - 1\right]
 + \frac{\pi d}{4\Omega}(n_+ L_+ + n_- L_-)^2 -\frac{\Omega \kappa_{eff}^3}{12 \pi} .
 \end{eqnarray}
Similarly, consider the limit of long rods, i.e. $L_k$ being large and approaching $\infty$. Approaching this limit 
while keeping the charge per rod fixed, the free energy becomes
 \begin{eqnarray}
 \frac{F^{iso}\left\{L_k \rightarrow \infty \right\}}{k_BT}  &=& \sum_{\gamma = +,-,c+,c-} n_\gamma \left[\ln \frac{n_\gamma}{\Omega} - 1\right]
  + \frac{\pi d}{4\Omega}(n_+ L_+ + n_- L_-)^2 + \frac{F_e^{iso}\left\{L_k \rightarrow \infty \right\}}{k_BT},\nonumber \\
  &&
 \end{eqnarray}
where the electrostatic contribution to the free energy is given by
 \begin{eqnarray}
 \frac{F_e^{iso}\left\{L_k \rightarrow \infty \right\}}{k_BT}  &=& \frac{\Omega}{2}\int \frac{d^{3}\mathbf{q}}{(2\pi)^3} \left\{\ln \left[1 + \frac{\kappa^2 }{q^2} + \frac{\pi \kappa_p^2}{q^3}\right] - \frac{\kappa^2
}{q^2} - \frac{\pi \kappa_p^2}{q^3}\right\}. \label{eq:fe_limit}
 \end{eqnarray}
Here, we have introduced  $\kappa_p^2 = \left(\frac{\kappa_+^2}{L_+}
+ \frac{\kappa_-^2}{L_-}\right)$. Furthermore, note that in writing Eq. ~\ref{eq:fe_limit}, we have used the aysmptotic form for $t^{iso}$ in the limit of $qL \rightarrow \infty$ for the 
entire range of $q$ (e.g., even in the case of $q = 0$). This causes the integral in Eq. ~\ref{eq:fe_limit} to 
diverge, while the original integral in Eq. ~\ref{eq:feiso} is convergent. These divergences are mere artifacts of the approximation scheme. Despite these divergences, we show that the leading contribution to the free energy of the rods is of the form $- n \sigma^2 l_B L\ln n/\Omega$, as described using the scaling arguments. 

For weakly charged polyelectrolytes so that $\kappa_p^2 \ll 1$, Eq. ~\ref{eq:fe_limit} can be 
written as (see Appendix B)
 \begin{eqnarray}
 \frac{F_e^{iso}\left\{L_k \rightarrow \infty \right\}}{k_BT}
 &\simeq&  -\frac{\Omega \kappa^3}{12 \pi}
 - \frac{\Omega \kappa_p^2}{4 \pi}\ln \left[\frac{\kappa}{q}\right]_{q\rightarrow 0} - \frac{\Omega \pi \kappa_p^4}{32 \kappa^3} + \cdots, \label{eq:fe_limit1} 
 \end{eqnarray}
where the leading term in the free energy expression has the functional form similar to the one 
described in Eq. ~\ref{eq:cyl_energy}. On the other hand, for strongly charged rods in weakly screened solutions $\kappa_p^2 \gg 1$ and $\kappa \rightarrow 0$. In this limit, $F_e^{iso}$ becomes (see Appendix B)
 \begin{eqnarray}
 \frac{F_e^{iso}\left\{L_k \rightarrow \infty \right\}}{k_BT}   &\simeq&  -\frac{\Omega \kappa^3}{12 \pi} - \frac{\Omega \kappa_p^2}{6 \pi} \ln \left[\frac{\sqrt{\pi}\kappa_p}{q}\right]_{q\rightarrow 0}  + \frac{\Omega \kappa^2 \kappa_p^{2/3}}{6\sqrt{3}\pi^{2/3}} \nonumber \\
 &&  - \frac{\Omega \kappa^4}{6\sqrt{3}\pi^{4/3}\kappa_p^{2/3}}- \frac{\Omega \kappa^2}{4 \pi^2} \left[q\right]_{q\rightarrow \infty} + \cdots. \label{eq:fe_limit2} 
  \end{eqnarray}
Like the weakly charged rods, the free energy of the strongly charged system has the same functional form as in Eq. ~\ref{eq:cyl_energy}. 

Using these approximate expressions for the free energy, the osmotic pressure can be readily computed using the thermodynamic relation $ P =  - \left(\partial F^{iso}/\partial \Omega\right)_{n,T}$. For the limiting case of $L_k \rightarrow 0$, this gives
 \begin{eqnarray}
 \frac{P\left\{L_k \rightarrow 0 \right\}}{k_BT}  &=& \sum_{k = +,-} \frac{\rho_k}{N_k} + \sum_{\gamma = c+,c-} \rho_\gamma -\frac{\kappa_{eff}^3}{24 \pi}, \label{eq:pressure_l0}
 \end{eqnarray}
which is the well-known Debye-H\"{u}ckel limiting law for
electrolyte solutions. Here, $\rho_k = n_k N_k/\Omega$ is the
monomer number density of type $k = +,-$ and $\rho_\gamma =  n_\gamma/\Omega$ is the number 
density of counterions of type $\gamma = c+,c-$. 

For the other limiting cases presented above, i.e., for $\kappa_p^2 \ll 1$
 \begin{eqnarray}
 \frac{P\left\{L_k \rightarrow \infty \right\}}{k_BT}  &=&  \sum_{k = +,-} \frac{\rho_k}{N_k} + \sum_{\gamma = c+,c-} \rho_\gamma -\frac{\kappa^3}{24 \pi} + \frac{\pi d}{4} (\rho_+ l_+ + \rho_- l_-)^2 - \frac{\kappa_p^2}{8 \pi} + \frac{\pi \kappa_p^4}{64 \kappa^3},
 \end{eqnarray}
 where $l_k \equiv L_k/N_k$ is the monomer length. Similarly, for $\kappa_p^2 \gg 1$ and $\kappa \rightarrow 0$
 \begin{eqnarray}
 \frac{P\left\{L_k \rightarrow \infty \right\}}{k_BT}  &=& \sum_{k = +,-} \frac{\rho_k}{N_k} + \sum_{\gamma = c+,c-} \rho_\gamma -\frac{\kappa^3}{24 \pi} + \frac{\pi d}{4}(\rho_+ l_+ + \rho_- l_-)^2 - \frac{\kappa_p^2}{12 \pi}  \nonumber \\
 && + \frac{\kappa^2 \kappa_p^{2/3}}{18 \sqrt{3}\pi^{2/3}} - \frac{\kappa^4}{9\sqrt{3}\pi^{4/3}\kappa_p^{2/3}}.
 \end{eqnarray}
Note that the leading contribution to the pressure coming from the
electrostatic correlations of the charged rods is negative and of
the form $-\kappa_p^2$ as already anticipated from the simple
scaling arguments in section ~\ref{sec:scaling_section}. This particular contribution
can drive  macrophase separation in isotropic solutions of rodlike
polyelectrolytes, a theme that is considered later in the paper.
Next, we turn to consider the weakly ordered nematic phase using the
approximate free energy given in Eq. ~\ref{eq:free_gen}.

\subsection{Weakly Ordered Nematic Phase}
For a nematic phase, the distribution function can be reduced to a
function of the cosine of the angle between the director and rod
orientation, i.e. $p_{k}(\mathbf{u},\nh) = p_{k}(\mathbf{u}\cdot
\nh)$ for $k=+,-$. In order to study a fully formed nematic phase,
the complete functional form for the probability distribution
function is needed. However, a stability analysis for a weakly
ordered nematic phase can be carried out without knowing the
probability distribution function \textit{a priori} by assuming a
two-term Legendre expansion for $p_{k}(\mathbf{u}\cdot \nh)$ written
as
 \begin{eqnarray}
 p_{k}(\mathbf{u}\cdot \nh) &\simeq& \frac{1}{4\pi}\left[1 + 5 S_k \frac{3(\mathbf{u}\cdot \nh)^2 - 1}{2}\right]. \label{eq:legend_expansion}
 \end{eqnarray}
 Here, $S_k$ is the nematic order parameter, given by
 \begin{eqnarray}
 S_k &=& \int d\mathbf{u} \: p_{k}(\mathbf{u}\cdot \nh) \left[ \frac{3(\mathbf{u}\cdot \nh)^2 - 1}{2} \right] .
 \end{eqnarray}
Here, we have used the normalization condition $\int d\mathbf{u} \:
p_{k}(\mathbf{u}\cdot \nh)  = 1$ to write the Legendre expansion.
Note that $S_k = 0$ for the isotropic phase and $1$ for the
completely ordered nematic phase, for which $p_{k}(\mathbf{u}\cdot
\nh) = \delta(\mathbf{u}-\nh)$. A similar analysis for the case of single component 
rodlike polyelectrolyte solutions has been carried out in Refs. \cite{potemkin_1,potemkin_2,potemkin_3}.

Using Eq. ~\ref{eq:legend_expansion}, $t_k(qL_k)$ for a weakly ordered
nematic phase can be written as $t_k(qL_k) = t^{iso}(qL_k) + \delta
t_k(qL_k)$, where $\delta t_k(qL_k)$ is given by
 \begin{eqnarray}
 \delta t_k(qL_k) &=& \frac{5 S_k}{4\pi}\int d\mathbf{u} \left[\frac{3(\mathbf{u}\cdot \nh)^2 - 1}{2}\right] \left[\frac{\sin(\mathbf{q}\cdot \mathbf{u} L_k/2)}{(\mathbf{q}\cdot \mathbf{u} L_k/2)}\right]^2 \\
 &=& \frac{15 S_k}{2} \left[3\frac{(\mathbf{q}\cdot \nh)^2}{q^2} - 1\right] \left[ \frac{1}{q^2L_k^2}\left(1 - \frac{\sin qL_k}{qL_k}\right) - \frac{t^{iso}(qL_k)}{6}\right]
 \end{eqnarray}
Using these expressions for $p_k$ and $t_k(qL_k)$, the free energy of
a weakly ordered nematic phase (expanded to second order in $S_k$)
can be written as
 \begin{eqnarray}
F & = & F^{iso} + \Delta F,
\end{eqnarray}
where
\begin{eqnarray}
\Delta F & = & \Delta F_{en} + \Delta F_{w} + \Delta F_{e}
\end{eqnarray}
so that
\begin{eqnarray}
\Delta F_{en} &=& \sum_{k=+,-}\frac{5}{2}n_k S_k^2, \\
\Delta F_w &=& -\frac{5\pi d}{32\Omega}\left(n_+ L_+ S_+ + n_- L_- S_-\right)^2, \\
 \Delta F_e   &=& - \frac{90\Omega}{\left(4 \pi\right)^2}\int_{0}^{\infty} dq q^2\frac{\left[\sum_{k=+,-} \kappa_k^2 S_k h(qL_k)\right]^2}{\left[q^2 + \kappa^2 + \sum_{k=+,-}\kappa_k^2 t^{iso}(qL_k)\right]^2}.
\end{eqnarray}
Here, we have used the approximation $\sin \theta \simeq \left[1 -
5P_2(\cos\theta)/8\right]\pi/4$, $P_2$ being the Legendre polynomial
of order $2$, in writing $\Delta F_w$, and
 \begin{eqnarray}
 h(qL_k) &=& \frac{1}{q^2 L_k^2}\left( 1 - \frac{\sin qL_k}{qL_k}\right) - \frac{t^{iso}(qL_k)}{6}.
 \end{eqnarray}
Note that the contribution to the free energy coming from
electrostatic correlations (i.e., $\Delta F_e$) is negative. In
other words, electrostatic correlations favor the nematic phase.
This observation will prove important to the the stability of the
weakly ordered nematic phase with respect to the isotropic phase.

Some useful results can be inferred from the above expression for
the free energy of a weakly ordered nematic phase. The spinodal of
the isotropic-nematic transition (i.e. stability limit of the
isotropic phase in the nematic region) is given by the condition
\begin{eqnarray}
\det \left[\begin{array}{cc}
\frac{\partial^2 F}{\partial S_+ \partial S_+} & \frac{\partial^2 F}{\partial S_+ \partial S_-}     \\
\frac{\partial^2 F}{\partial S_+ \partial S_-} & \frac{\partial^2 F}{\partial S_- \partial S_-}
\end{array} \right ]  &=& 0 \label{eq:spino_iso_nem}
\end{eqnarray}
For the case of symmetric mixtures, i.e. equal length ($=L$),
diameter ($=d$), and number density of the rods ($=n/\Omega$), Eq.
~\ref{eq:spino_iso_nem} reduces to
\begin{eqnarray} 
S_{IN} = \frac{25 n^2 }{4 }\left[1 - \frac{\pi}{16}c_p   - \left(4 \pi \tilde l_B Z^2 \sigma^2 N\right)^2 \Gamma\left\{c_p\right\} \frac{L}{d}c_p \right]  &=& 0 \label{eq:iso_nem_spino}
\end{eqnarray}
where
 \begin{eqnarray}
 \Gamma\left\{c_p\right\} &=& \frac{9}{4 \pi^2 }\int_{0}^{\infty} d\tilde{q} \tilde{q}^2\frac{h^2(\tilde{q})}{\left [\tilde{q}^2 + \tilde \kappa^2 + \tilde \kappa_{sm}^2 t^{iso}(\tilde{q})\right ]^2},
 \end{eqnarray}
 and $c_p = 2nL^2 d/\Omega$ is the \textit{dimensionless} parameter of the order of 
 overlap concentration of the rods. For numerical purposes, dimensionless 
 parameters are introduced after renormalizing different parameters with the length of the rods so that 
 $\tilde \kappa^2 = \kappa^2 L^2 = 4\pi \tilde l_B \sigma c_p L l/d, \; \tilde
\kappa_{sm}^2 = L^2(\kappa_+^2 + \kappa_-^2) = 4\pi \tilde l_B
\sigma^2 c_p L^2 l/d$, and $\tilde q = qL$. Furthermore, in order to make a qualitative 
connection with the Manning's theory of counterion condensation\cite{manning} later in this work, we 
have renormalized Bjerrum's length using monomer length ($l \equiv L/N$) and defined $\tilde l_B = l_B/l$. 

Note that for the symmetric mixtures, the spinodal can also be
obtained after putting $S_+ = S_- = S$ and then evaluating
$\partial^2 F/\partial S^2 = 0$. Furthermore, $S_{IN}<0$ corresponds
to the  regime where the isotropic phase is unstable relative to the
weakly ordered nematic phase. It is clear from Eq.
~\ref{eq:iso_nem_spino} that the electrostatics and the steric
effects (last two terms in square brackets) act against the
orientational entropy to drive the system from the isotropic phase
to the anisotropic nematic phase. In the absence of
electrostatics\cite{onsager_paper,edwardsbook}, the nematic phase
becomes stable when $c_p > 16/\pi \simeq 5.1$. It is also clear from
Eq. ~\ref{eq:iso_nem_spino} that the nematic phase in
polyelectrolyte mixtures becomes stable at lower polymer
concentrations compared to the neutral mixtures.

\begin{figure}[ht!]
\vspace*{-0.95cm}
\begin{center}
$\begin{array}{c@{\hspace{.001in}}c@{\hspace{.001in}}}
\includegraphics[width=3.5in,height=3.5in]{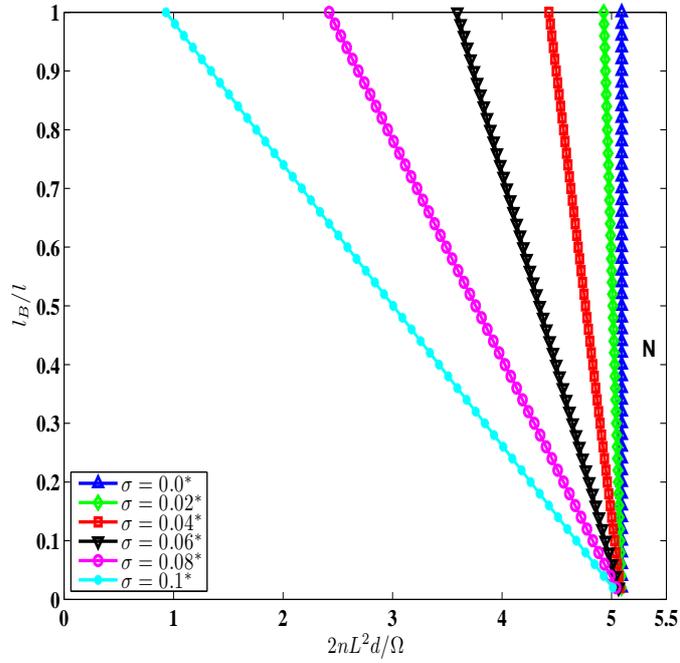}&   \\
\textbf{(a)} & \\
& \\
\includegraphics[width=3.5in,height=3.5in]{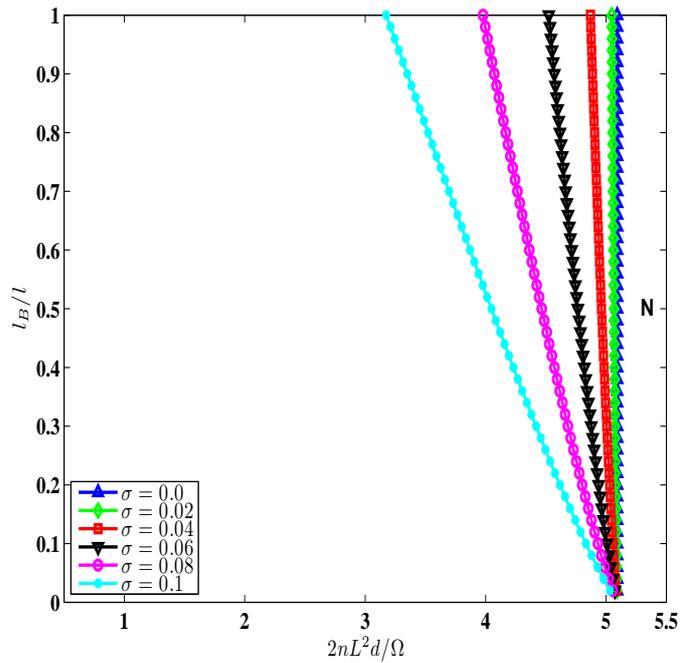} & \\
\textbf{(b)} &
\end{array}$
\vspace*{-1.0cm}
\newline
\end{center}
\caption{Phase boundaries (spinodals) for the stability of a weakly
ordered nematic phase for a symmetric mixture of rodlike
polyelectrolytes in the absence (a) and the presence of the
counterions (b). On the right hand side of each boundary, the
nematic phase (denoted by ``N'') is stable in comparison with the
isotropic phase. These results are obtained for $N = 1000, \; d/l =
1/50$.} \label{fig:compare_spinod_in_rods}
\end{figure}

Numerical solutions of Eq. ~\ref{eq:iso_nem_spino} are presented in
Fig. ~\ref{fig:compare_spinod_in_rods} for  symmetric mixtures
without and with the counterions (Fig.
~\ref{fig:compare_spinod_in_rods}a and
~\ref{fig:compare_spinod_in_rods}b, respectively) and for different
linear charge densities. On the right hand side of these boundaries,
the nematic phase is stable. It is clear from Fig.
~\ref{fig:compare_spinod_in_rods} that the nematic phase becomes
stable at lower monomer densities with an increase in linear charge
density at fixed Bjerrum length. Furthermore, comparing Fig.
~\ref{fig:compare_spinod_in_rods}a and
~\ref{fig:compare_spinod_in_rods}b, it is evident that the nematic
phase becomes stable at higher monomer densities in the presence of
counterions, i.e. the stable regime of the nematic phase is smaller
in the presence of counterions. Note that these numerical results
are in agreement with earlier theoretical predictions for
one-component rodlike polyelectrolyte solutions that electrostatic
interactions favor uniaxial ordering of the rods.\cite{potemkin_1,potemkin_2,potemkin_3}

\subsection{Nematic Phase for Symmetric Mixtures: Variational Treatment}
In order to study a nematic phase with an arbitrary magnitude of the
orientational order parameter, we need to resolve the full
probability distribution function. In principle, this can be done by
a calculus of variations approach by minimizing the free energy of
the system, which leads to an integral equation\cite{edwardsbook}.
For the case of neutral polymers, the problem has been attacked by
three different routes. The first is a variational treatment using
Onsager's trial
function\cite{edwardsbook,onsager_paper,onsager_trial_func}, where
the variational parameter is determined by minimizing the free
energy. A second route is through the use of a Legendre
expansion\cite{lakatos_legendre,lasher_legendre} and determining the
coefficients in the Legendre series that minimize the free energy. A
third scheme is to directly attack the integral equation in real
space\cite{intgral_equation_real} using a non-linear equation
solver. From a computational point of view, the last two routes are
more demanding and become especially difficult for strongly ordered
nematic phases ($S>0.9$). The first route is the easiest and readily
describes a nematic phase with arbitrary order. However, it leads to
a slight overprediction\cite{edwardsbook,lasher_legendre} of the
coexisting densities at the isotropic-nematic transition.

In this work, we adopt Onsager's variational approach to study
nematic phases in the symmetric polyelectrolyte mixtures. Due to the
assumed symmetry, the probability distribution functions for the two
types of rods must be the same and only one of the distribution
functions needs to be considered. In Onsager's approach, the
probability distribution function is taken to be of the form
\begin{equation}
p(\mathbf{u}\cdot \nh) =  \frac{\alpha \cosh \left[\alpha
\left(\mathbf{u}\cdot \nh\right)\right]}{4 \pi \sinh \alpha},
\end{equation}
where $\alpha$ is a variational parameter, which is determined by
minimization of the free energy. The order parameter corresponding
to this distribution function for a symmetric mixture is
\begin{equation}
S = 1 + \frac{3}{\alpha^2}(1-\alpha \coth \alpha)
\end{equation}
so that $S=0$ for $\alpha = 0$ corresponds to the isotropic phase
and $S=1$ when $\alpha \rightarrow \infty$, corresponding to a
perfectly ordered nematic phase. The free energy density can be
written in terms of $\alpha$ so that the free energy density of the
symmetric mixtures becomes $f =  F/\Omega = f_{en} + f_{w} + f_e $,
where

\begin{eqnarray}
   \frac{f_{en}L^2 d}{k_BT} & = & c_{p}\left[\ln \left[\alpha \coth \alpha \right] - 1 + \frac{\tan^{-1}\left[\sinh \alpha\right]}{\sinh\alpha}\right] + c_{s} \left[\ln \frac{c_s}{2L^2d} - 1\right] \nonumber \\
   && + c_{p} \left[\ln \frac{c_p}{2L^2d} - 1\right], \label{eq:free_en_nem}\\
 \frac{f_w L^2 d}{k_BT} &=&  \frac{\pi}{2} \frac{I_2(2\alpha)}{\left[\sinh\alpha\right]^2}c_p^2,\label{eq:free_w_nem}\\
\frac{f_e L^2 d}{k_BT} &=&   \frac{d}{8\pi^2
L}\int_{0}^{\infty}d\tilde{q}\: \tilde{q}^2\int_{-1}^{1}
dx\left\{\ln \left[1 + \frac{\tilde{\kappa}^2 + \tilde\kappa_p^2
t(\tilde{q},x)}{\tilde{q}^2}\right] -  \frac{\tilde{\kappa}^2 +
\tilde \kappa_p^2
t(\tilde{q},x)}{\tilde{q}^2}\right\}\label{eq:free_rods_2}
\end{eqnarray}
and
 \begin{eqnarray}
    t(\tilde{q},x)  &=& \int_{-1}^{1}dx' \: \frac{\alpha \cosh \left[\alpha x'\right]}{2 \pi \sinh \alpha}\int_{-1}^{1}\frac{dy}{\sqrt{1-y^2}}\left[\frac{\sin\left[\tilde{q}\left(xx' + \sqrt{1-x^2}\sqrt{1-x'^2} \:y\right)/2\right]}{\left[\tilde{q}\left(xx' + \sqrt{1-x^2}\sqrt{1-x'^2} \: y\right)/2\right]}\right]^2 \label{eq:tfunc_var_nem}
\end{eqnarray}
where $c_s = \sigma L c_p$ is the dimensionless number density of
small ions (both positive and negative) and $I_2$ is the modified
Bessel function of order $2$. The integrals in the expression for
$t(\tilde{q},x)$ can be readily evaluated using Gauss-Legendre and
Gauss-Chebyshev quadratures for $x'$ and $y$ integrals,
respectively. Also, the integral over $\tilde{q}$ ranging from $0$
to $\infty$ can be evaluated using Gauss-Legendre quadrature after
using the transformation $q = (1-z)/(1+z)$.

\begin{figure}[ht!]
\vspace*{-0.95cm}
\begin{center}
$\begin{array}{c@{\hspace{.001in}}c@{\hspace{.001in}}}
\includegraphics[width=3.5in,height=3.5in]{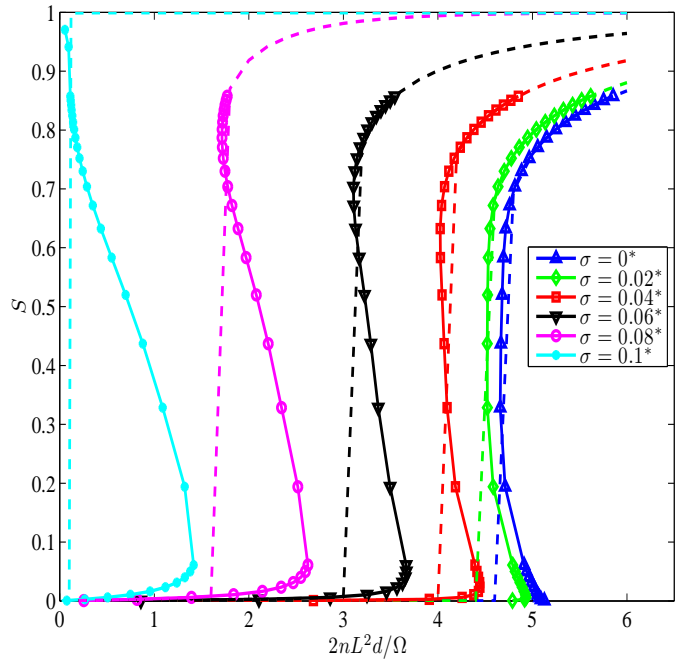}&   \\
\textbf{(a)} & \\
& \\
\includegraphics[width=3.5in,height=3.5in]{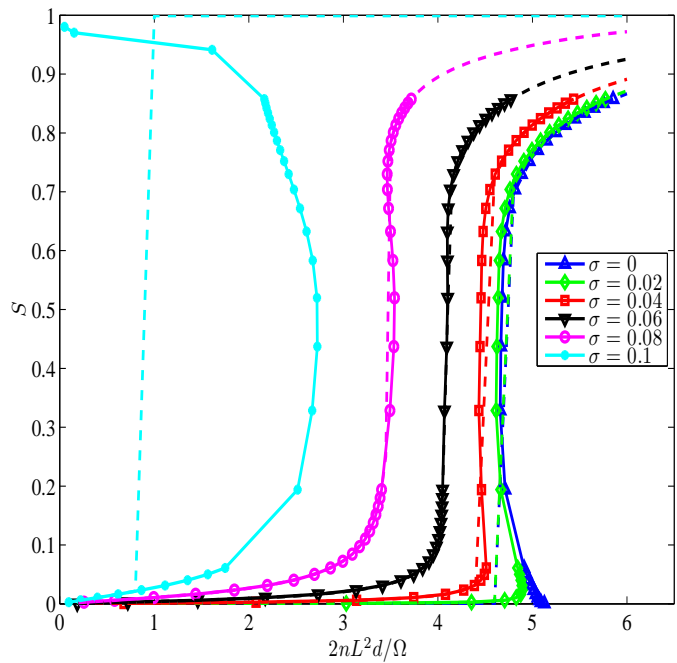} & \\
\textbf{(b)} &
\end{array}$
\vspace*{-1.0cm}
\newline
\end{center}
\caption{Effect of the linear charge density $\sigma$ on the
isotropic-nematic phase transition (characterized by jump in order
parameter) in salt-free symmetric mixtures of oppositely charged
rodlike polyelectrolytes. Figs. (a) and (b) correspond to trends in
the nematic order parameter $S$ with respect to dimensionless rod
concentration $2 n L^2 d/\Omega$ for symmetric mixtures without and
with counterions, respectively. In each figure, the solid lines
present the solutions obtained by solving the nonlinear equations
for the optimal variational parameter in Onsager's trial function
and the dashed lines are the results of numerical minimization of
the free energy with respect to the same variational parameter. The
curves correspond to $N=1000, \; d/l = 1/50$ and $l_B/l = 0.7$. }
\label{fig:in_bet_n1000}
\end{figure}

The free energy density can be optimized with respect to $\alpha$ so that $\frac{\partial f}{\partial \alpha} = 0$, which gives
\begin{eqnarray}
c_{p}\left[\frac{1}{\alpha} - \frac{\tan^{-1}\left[\sinh \alpha\right]\cosh \alpha}{\left[\sinh\alpha\right]^2}\right] + \frac{\pi c_p^2}{2\left[\sinh\alpha\right]^2} \left[I_1(2\alpha) + I_3(2\alpha) - 2\coth\alpha I_2(2\alpha)\right]&&\nonumber\\
- \frac{d}{8\pi^2 L}\int_{0}^{\infty}d\tilde{q} \int_{-1}^{1} dx  \frac{\partial t(\tilde{q},x)}{\partial \alpha}\left\{\frac{\tilde \kappa_p^2\left[\tilde \kappa^2
+ \tilde \kappa_p^2 t(\tilde{q},x)\right]}{ \tilde{q}^2 + \tilde \kappa^2
+ \tilde \kappa_p^2 t(\tilde{q},x)}\right \} = 0 \qquad \qquad \qquad && \label{eq:order_equation}
\end{eqnarray}
Note that for a root ($=\alpha^\star$) of
Eq.~\ref{eq:order_equation} to be a minimum of the free energy,
$\frac{\partial^2 f}{\partial \alpha^2}|_{\alpha =  \alpha^\star} >
0 $ must be satisfied. In other words, some of the roots of Eq.
~\ref{eq:order_equation} may correspond to a local maximum in the
free energy density rather than a local minimum. To ensure that we
retain only the physical roots, we have conducted tandem numerical
solutions of Eq.~\ref{eq:order_equation} and direct minimization of
the free energy. In the former, instead of solving for $\alpha$
using different values of $c_p$, it proves easier to solve
Eq.~\ref{eq:order_equation} for $c_p$ using different values of
$\alpha$. In second set of calculations, we have carried out
numerical minimization (using Brent's method\cite{numerical_recipe})
of the free energy with respect to $\alpha$ for different values of
$c_p$. Results of these two sets of calculations are presented in
Fig.~\ref{fig:in_bet_n1000}. Fig.~\ref{fig:in_bet_n1000}a presents
the results of the calculations without counterions and Fig.
~\ref{fig:in_bet_n1000}b corresponds to results with counterions.
Solid and dashed lines correspond to the roots of
Eq.~\ref{eq:order_equation} and the numerical minimization,
respectively.

In both the figures, the results of the two sets of calculations
match perfectly well except in the transition regime, where there
are multiple values of the order parameter $S$ for a given value of
$c_p$. In fact, this is the metastable regime for the
isotropic-nematic phase transition and such a diagram has already
been mapped out for the neutral rods using bifurcation
analysis\cite{bifurcation}. The jump in $S$ for the numerical
minimization calculations (i.e., dashed lines) is a characteristic
of the first order isotropic-nematic phase transition. From both the
figures, it is clear that the isotropic-nematic phase transition
takes place at lower monomer densities as the linear charge
densities of the rodlike polyelectrolytes is increased. Also, in
comparing the two figures, it is clear that the isotropic-nematic
phase transition takes place at higher monomer densities in the
presence of counterions. These results are consistent with the
stability analysis of the weakly ordered nematic phase as presented
in the previous section.

We note that the isotropic-nematic transition in this solvated
system is actually spanned by a region of two-phase coexistence in
which a  diluted (in polymer) isotropic phase coexists with an
enriched nematic phase. Such two-phase regions are confined to
within the regions of hysteresis shown in
Fig.~\ref{fig:in_bet_n1000}, but we have not taken the trouble to
elaborate them in this figure. The two-phase regions will be
revealed when we consider the full phase diagram. 

However, before getting to the full phase diagram, an important remark on the 
theoretical treatment considered in this paper is appropriate. During the 
computation of the full phase diagram, it is found that for some set of parameters 
a completely ordered nematic phase (i.e., the phase for which $\alpha \rightarrow \infty$ and hence, $S = 1$) 
becomes one of the competing phases. Physically, the completely ordered phase corresponds to 
perfectly aligned rods. In the next section, we show that Onsager's approach (which is extended to polyelectrolyte mixtures in this work) is not able to describe such a coexistence between the completely ordered nematic 
phase and any other phase.    

\subsection{Completely Ordered Nematic Phase for Symmetric Mixtures}
For the completely ordered nematic phase, Onsager's variational parameter, $\alpha$, 
diverges. In order to evaluate the free energy in this limit, we rewrite Eqs. ~\ref{eq:free_en_nem}-~\ref{eq:free_w_nem} using  
asymptotic expansions for the Bessel and hyperbolic functions\cite{onsager_paper} leading to the result 
\begin{eqnarray}
   \frac{f_{en}^{h}L^2 d}{k_BT} & = & c_{p}\left[\ln \alpha  - 1 \right] + c_{s} \left[\ln \frac{c_s}{2L^2d} - 1\right] + c_{p} \left[\ln \frac{c_p}{2L^2d} - 1\right], \label{eq:free_c_en_nem}\\
 \frac{f_w^h L^2 d}{k_BT} &=&  \sqrt{\frac{\pi}{\alpha}} c_p^2\left[1 - \frac{30}{32 \alpha} + \frac{210}{\left(32 \alpha\right)^2} + \frac{1260}{\left(32 \alpha\right)^3} + \cdots \right].\label{eq:free_c_w_nem}
\end{eqnarray}
In the limit of $\alpha \rightarrow \infty$, the entropic contribution to the 
free energy of the completely ordered nematic phase diverges logarithmically (i.e., $f_{en}^c =  f_{en}^{h}\left\{\alpha \rightarrow \infty \right \}\rightarrow \infty$) 
and the excluded volume contribution vanishes (i.e., $f_{w}^c =  f_{w}^{h}\left\{\alpha \rightarrow \infty \right \} \rightarrow 0$). Furthermore, the limiting expression for Eq. ~\ref{eq:tfunc_var_nem} becomes 
 \begin{eqnarray}
     t^c(\tilde{q}x) &\equiv& t(\tilde{q},x)|_{\alpha \rightarrow \infty}  = \left[\frac{\sin\left[\tilde{q}x/2\right]}{\tilde{q}x/2}\right]^2, \label{eq:tfunc_c_var_nem}
\end{eqnarray}
which allows us to rewrite Eq. ~\ref{eq:free_rods_2} as 
\begin{eqnarray} 
\frac{f_e^c L^2 d}{k_BT} &=&   \frac{d}{\pi^2
L}\int_{0}^{\infty}d\tilde{q}\: \tilde{q}^2\left\{\frac{\tilde{\kappa}^2 + \tilde\kappa_p^2
t^c(2\tilde{q})}{4\tilde{q}^2} - \left[1 + \frac{\tilde{\kappa}^2 + \tilde\kappa_p^2
t^c(2\tilde{q})}{4\tilde{q}^2}\right]\ln \left[1 + \frac{\tilde{\kappa}^2 + \tilde\kappa_p^2
t^c(2\tilde{q})}{4\tilde{q}^2}\right] \right\}.\nonumber \\
&&\label{eq:free_rods_2_c}
\end{eqnarray}
This integral can be evaluated numerically and is found to be negative. Note that Eq. ~\ref{eq:free_rods_2_c} 
is the same as the electrostatic contribution to the free energy considered in Ref. \cite{stell_rods} 
in the context of phase separation of charged aligned needles. 

The divergence of $f_{en}^c$ complicates the evaluation of coexisting densities between the isotropic and the completely ordered phases. Also, the vanishing of $f_w^c$ for \textit{any} 
arbitrary polymer density highlights the limitation of the Onsager's virial approach to describe phase separation when one of the competing phases is fully ordered. 

A more suitable description of such highly ordered phases is presented in Ref. \cite{stell_rods}, where the phase separation of a completely ordered parent phase into two 
completely ordered daughter phases having different polymer densities is considered. The phase separation calculations were performed using the Percus-Yevick equation of state for hard cylinders. This particular description takes into account the higher order terms in describing the excluded volume interactions and goes beyond the Onsager approach. However, this study did not consider the nematic phase as a candidate in the free energy competition. Such an analysis, while possibly relevant in an orientationally constrained situation, could produce unphysical results if the system can freely choose the orientation and concentration of both parent and daughter phases. 

In the following, in order to avoid the above complications associated with strongly ordered nematic phases, we 
limit our results to regimes where the nematic order parameter is not fully saturated at $S=1$.

\section{Results} \label{sec:results}

Using the theoretical approach and free energies presented in the
previous section, we have investigated the phase behavior of
symmetric mixtures of oppositely charged rodlike polyelectrolytes.
In the following, we consider the possibility of macrophase
separation in the isotropic phase due to attractive electrostatic
correlations between oppositely charged polyelectrolytes in addition
to phase coexistence between isotropic and the nematic phases.
Unless mentioned, all lengths are normalized by the monomer length
$l$ so that $L = Nl$, with $N$ the number of monomers.  In order to
keep the theoretical analysis simple and avoid the issue of
counterion condensation\cite{manning}, we have focused on the
``weakly charged'' regime corresponding to $l_B/l < 1$. Also, we
restrict attention to salt-free symmetric mixtures here and leave
the effect of added salt on the phase behavior for the future. In
order to identify the role played by the counterions in the
salt-free symmetric mixtures, we consider two comparison systems -
one without any counterions from the polyelectrolytes and the other
with counterions. As noted above, the counterion free situation
could conceptually be realized in symmetric mixtures containing
polyacids and polybases, where the charge on the polyelectrolyte
backbones can be controlled by the pH of the solution. While the
theory is quite general and a wide variety of phenomena can be
investigated, we have chosen to further limit the parameter space of
our study by fixing the rod length and aspect ratio to $N = 1000$
and $d/l = 1/50$.

Prior to presenting results on the full phase diagram, it is
illustrative to study macrophase separation in the isotropic phase,
where we set aside the possibility of nematic order. Even for such a
simpler situation, there are some key questions that need to be
answered. For example, it is not clear how the counterions are
partitioned between different phases and where the phase boundaries
(spinodals and binodals) are located relative to the analogous phase
boundaries in mixtures containing flexible polyelectrolytes. In the
next subsection, we present our results for phase separation in the
isotropic phase and address these key issues. In a subsequent
subsection, we present the full phase diagram by considering the
possibility of isotropic-nematic phase behavior in addition to the
isotropic-isotropic transition.

\subsection{Isotropic-isotropic transition}
A salt-free symmetric mixture of oppositely charged polyelectrolytes
with counterions is a system with five components - two types of
polymers, two types of counterions and the solvent molecules.
However, in the simplified theoretical description presented here,
the solvent molecules are treated implicitly and by restricting
attention to symmetric mixtures only, the equations obtained by
enforcing the equality of the chemical potentials of the polymers in
the two phases are degenerate for the two types of polymers. The
same is true for the counterions. As a consequence, we only have to
analyze a pseudo two-component system, where the polymers and the
counterions need to be partitioned among the coexisting phases.
Using the same set of arguments, salt-free symmetric mixtures
without counterions can be treated as a pseudo one-component system.

\subsubsection{Effect of counterions }
For the quasi-one and two component systems, we have carried out a
direct minimization of the total free energy density, obtained by
appropriate weighting of two isotropic phases and the use of the
lever rule\cite{chandler_book}. This approach is equivalent to
equating the chemical potential of each component in the two phases
and equating the pressure of each phase, but replaces solving a
nonlinear system by the numerically more robust procedure of
minimization. Explicitly, for the quasi-one component system, we
minimize the total free energy density ($f_{total}^{iso}$) of the
phase segregated system with respect to the densities in each phase
(i.e., two dimensional minimization) after writing
\begin{eqnarray}
f_{total}^{iso} &=& \frac{c_{p}^{\text{II}} - c_p}{c_{p}^{\text{II}} - c_{p}^{\text{I}}}f^{iso}\left\{c_{p}^{\text{I}}\right\} + \frac{c_p - c_{p}^{\text{I}}}{c_{p}^{\text{II}} - c_{p}^{\text{I}}}f^{iso}\left\{c_{p}^{\text{II}}\right\}, \label{eq:ftotal_iso}
\end{eqnarray}
where $f^{iso}\left\{c_{p}^{\text{I}}\right\}$ and
$f^{iso}\left\{c_{p}^{\text{II}}\right\}$ are the free energy
densities of isotropic phases $\text{I}$ and $\text{II}$,
respectively, given by Eq.~\ref{eq:free_iso}. The parameters
$c_p^{\text{I}}$, $c_p$, and $c_p^{\text{II}}$ ($c_{p}^{\text{I}} <
c_p < c_{p}^{\text{II}}$) are the dimensionless number density of
the polyelectrolytes in dilute phase $\text{I}$, overall number
density, and the number density in the concentrated phase
$\text{II}$, respectively. A similar equation can be written for the
quasi-two component system (i.e. the system with counterions) where
we note that the total number of counterions is related to the total
number of monomers by $n_c^{total} =  2 \sigma n_t L$, where $n_t$
is the total number of polyelectrolyte chains of one type. For this
system, three dimensional minimizations of the free energy with
respect to the monomer densities in each phase along with the
counterion density in one of the phases have been carried out to
compute the coexistence curves. The counterion density in the second
phase is computed using the lever rule for the counterions. All of
the multi-dimensional minimizations of the free energy density have
been carried out using the simplex method\cite{numerical_recipe}.

\begin{figure}[ht!]
\vspace*{-0.95cm}
\begin{center}
$\begin{array}{c@{\hspace{.001in}}c@{\hspace{.001in}}}
\includegraphics[width=3.5in,height=3.5in]{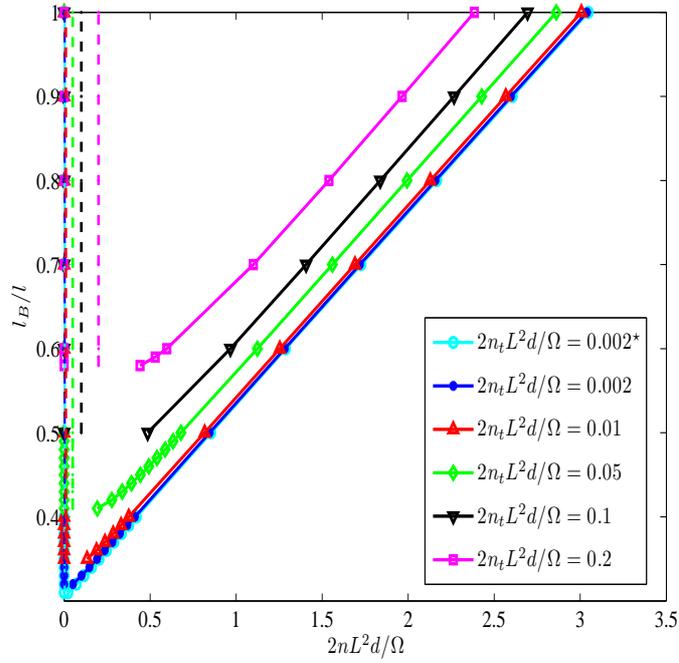}&   \\
\textbf{(a)} & \\
& \\
\includegraphics[width=3.5in,height=3.5in]{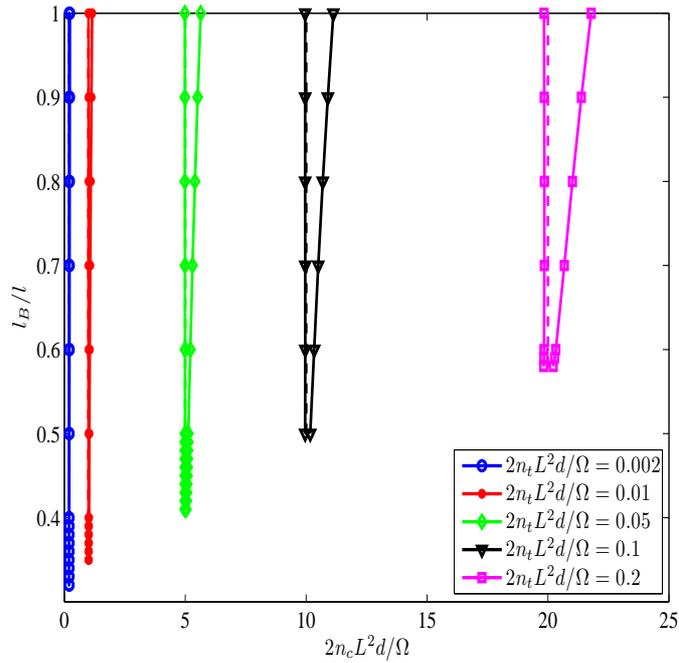} & \\
\textbf{(b)} &
\end{array}$
\vspace*{-1.0cm}
\newline
\end{center}
\caption{Coexistence curves (binodals) for isotropic-isotropic phase
separation in salt-free symmetric mixtures of oppositely charged
rodlike polyelectrolytes. Polyelectrolyte density and counterion
density in the coexisting phases are presented in Figs. (a) and (b),
respectively. $2n_t/\Omega$ corresponds to the total number density
of rodlike polyelectrolytes and these curves were obtained using
$N=1000,\; \tilde d = 1/50,\; \sigma = 0.1$. In Fig. (a), the plot
with $2n_t L^2 d/\Omega = 0.002^\star$ represents the binodal for a
symmetric mixture without any counterions for comparison purposes.
Also, in Fig. (b), the counterion density in the dense phase
corresponds to the right hand side of the dashed lines. In Figs. (a)
and (b), dashed lines indicate the total number densities of
polyelectrolytes and counterions, respectively.}
\label{fig:ii_bet_n1000}
\end{figure}

In Fig.~\ref{fig:ii_bet_n1000}, we present the results of these
calculations for different total monomer densities. Coexisting
monomer and counterion densities in the two phases are presented in
Figs.~\ref{fig:ii_bet_n1000}a and ~\ref{fig:ii_bet_n1000}b,
respectively. In these calculations, the linear charge density
$\sigma$ for the polyelectrolytes is kept the same so that different
total monomer number densities correspond to different total
counterion number densities. Also, for comparison purposes, the
coexisting monomer densities in the counterion-free system are also
presented in Fig.~\ref{fig:ii_bet_n1000}a for a total monomer
density of $0.002$, and are denoted by $\star$. Note that in
Fig.~\ref{fig:ii_bet_n1000}b the counterion densities on the right
side of the dashed lines correspond to the densities in the phase
with higher monomer density.

It should be noted that these results are in qualitative agreement
with similar calculations for flexible polyelectrolytes, where the
system phase separates into a very low density (supernatant) phase
and a dense (coacervate) phase. The asymmetric nature of the
coexistence curve\cite{fisher_levin,fisher} is a result of the
long-range electrostatic interactions at play in these systems
coupled with the reduced translational entropy of polymers relative
to solvent and ions. Beyond these observations, two important
results can be inferred from the figures. In
Fig.~\ref{fig:ii_bet_n1000}a, it is clear that the coexistence
regime shrinks continuously with the increase in the total
counterion density. This implies that the counterions suppress the
isotropic-isotropic phase transition. This is in agreement with the
notion that the counterions need to be partitioned among the phases
upon macrophase separation, which is entropically unfavorable.
Furthermore, from Fig. ~\ref{fig:ii_bet_n1000}b, it is found that
indeed, the counterions get partitioned between the two phases and
the counterion density is only slightly higher in the coacervate
phase. Note that this result is in qualitative agreement with the
theoretical and experimental results reported by
Voorn\cite{voorn_5}, although the treatment of electrostatics in
Voorn's theory is much more primitive than ours.

An important remark regarding the phase diagrams presented in this work
is due here. Typically, in a two-phase region 
the boundaries of the phase diagram are the same irrespective of the initial 
concentration; just the relative amounts
of the two phases vary. The situation in the presence of counterions is very different. 
A change in the initial concentration of polyelectrolytes changes the concentration 
of counterions in the solution and the phase boundaries may shift as described in Fig. ~\ref{fig:ii_bet_n1000}.   
As different concentrations of the polyelectrolytes and their counterions correspond to 
different states of the system, the diagrams presenting the coexisting phases for different initial 
states should be called ``state diagrams'' in general. However, in this work, we ignore this semantics issue and 
call these diagrams binodals (or coexistence curves). 

\subsubsection{Comparison with flexible polyelectrolytes: counterion free symmetric mixtures}

In order to compare the phase boundaries of symmetric mixtures
containing rodlike polyelectrolytes to those of flexible
polyelectrolyte mixtures, we have computed the free energy of
mixtures of flexible polyelectrolytes using an analogous random
phase approximation to that employed in the rigid case (see Appendix
C). Explicitly, the free energy of a flexible mixture is given by
\begin{eqnarray}
        F_f &=& F_{en}^f + F_{w}^f + F_{e}^f  \label{eq:free_flex}
        \end{eqnarray}
        where
        \begin{eqnarray}
       \frac{F_{en}^{f}}{k_BT} &=& -\sum_{k=+,-} n_k \ln Q_k^0 + \sum_{\gamma = +,-,c+,c-} n_{\gamma} \left[\ln \frac{n_{\gamma}}{\Omega} - 1\right] \\
       \frac{F_{w}^{f}}{k_BT} &=& \frac{w}{2 \Omega}\left( n_+ N_+ + n_- N_-\right)^2  \nonumber \\
&& + \frac{\Omega}{2}\int \frac{d^{3}\mathbf{q}}{(2\pi)^3} \left\{\ln \left[1 + \frac{w}{\Omega}\sum_{k=+,-}N_{k}^2 n_k g\left(q^2N_k l_k^2/6\right)\right]  - \frac{w}{\Omega}\sum_{k=+,-}N_{k}^2 n_k g\left(q^2N_k l_k^2/6\right)\right \}\nonumber \\
&& \\
\frac{F_{e}^{f}}{k_BT} &=& \frac{\Omega}{2}\int \frac{d^{3}\mathbf{q}}{(2\pi)^3} \left\{\ln \left[1 + \frac{\kappa^2
+ \sum_{k=+,-}\kappa_{k}^2 g\left(q^2N_k l_k^2/6\right)}{q^2}\right] - \frac{\kappa^2
+ \sum_{k=+,-}\kappa_{k}^2 g\left(q^2N_k l_k^2/6\right)}{q^2} \right \} \nonumber \\
&&
\end{eqnarray}
Here, $w$ is the excluded volume parameter\cite{edwardsbook} and
$Q_k^0$ is the partition function for a noninteracting Gaussian
chain of length $L_k = N_k l_k, k = +,-$ ($l_k$ being the Kuhn
segment length and $N_k$ is the number of
segments)\cite{edwardsbook}). Furthermore, $g(x) =  2 (e^{-x} - 1 +
x)/x^2$ is the Debye function. We note that in the case of flexible
polyelectrolyte mixtures the final term in $F_w^f$ (involving the
integral) is the well-known Edwards' screening contribution to the
free energy\cite{edwardsbook} and is negative. For the comparison
between the rodlike and the flexible systems, this contribution will
be ignored and its effect captured by using a renormalized excluded
volume parameter $w_r$ instead of the bare excluded volume parameter
$w$.

Since our focus here is on the effect of chain flexibility on the
electrostatic contribution to the free energy, we can identify an
appropriate value of $w_r$ by forcing agreement between the excluded
volume contributions to the free energy for the rodlike and flexible
symmetric mixtures (cf. Eqs.~\ref{eq:free_iso} and
\ref{eq:free_flex}). The comparison reveals that $w_r = \pi d l^2/2$
is the suitable choice, which makes all free energy contributions
other than electrostatics identical for the rodlike and flexible
systems. Using this value for the renormalized excluded volume
parameter, the spinodals and the binodals for symmetric mixtures of
flexible polyelectrolytes can be readily computed using the same
approach as used for the rodlike system. To avoid any complications
arising from the presence of counterions in comparing the phase
boundaries for symmetric rodlike and flexible mixtures, we have
considered a salt-free and counterion-free system.

For this model flexible polyelectrolyte mixture, the spinodal for
the isotropic-isotropic transition is given by
\begin{eqnarray}
S_f^{iso} =  1 + c_p \left[w_r - \left(4 \pi \tilde l_B Z^2 \sigma^2 N\right)^2 \frac{L}{d}D_e\left\{c_p\right\}\right]  &=& 0 \label{eq:iso_iso_spino}
\end{eqnarray}
where
 \begin{eqnarray}
 D_e\left\{c_p\right\} &=& \frac{1}{4 \pi^2 }\int_{0}^{\infty} d\tilde{q} \tilde{q}^2\frac{g^2\left(\tilde{q}^2/6N\right)}{\left [\tilde{q}^2 + \tilde \kappa_{sm}^2 g^2\left(\tilde{q}^2/6N\right)\right ]^2},\label{eq:iso_iso_de}
 \end{eqnarray}
and $c_p = 2nL^2 d/\Omega$ is the corresponding
\textit{dimensionless} monomer number density. Also, as before, we
have defined $\tilde l_B = l_B/l, \; \tilde \kappa = \kappa L$,
 $\tilde \kappa_{sm}^2 = L^2(\kappa_+^2  + \kappa_-^2)$, and $\tilde q =
qL$. The spinodal for the isotropic-isotropic phase transition in a
corresponding \emph{rodlike} symmetric polyelectrolyte mixture can
be obtained from Eqs. (\ref{eq:iso_iso_spino} - \ref{eq:iso_iso_de})
by replacing the Debye function $g\left(\tilde{q}^2/6 N\right)$ by
the function $t^{iso}(\tilde{q})$ given in Eq.~\ref{eq:tkq_iso}.

The set of parameters that lead to $S_f^{iso}<0$ corresponds to the
regime of instability to macrophase separation in the isotropic
phase. Note that the electrostatic term $D_e \left\{c_p\right\}$ is
positive and hence, the electrostatics drives the macrophase
separation in flexible as well as rodlike polyelectrolyte mixtures.
Also, from Eq.~\ref{eq:iso_iso_spino} it is clear that the
translational entropy (which appears as unity in the equation) and
the excluded volume interactions oppose this driving force (assuming
$w_r > 0$ for good solvents). Furthermore, we note the prefactor of
$\left(4 \pi \tilde l_B Z^2\sigma^2 N\right)^2$ in front of $D_e
\left\{c_p\right \}$. This implies that an increase in the polymer
length, Bjerrum length or the linear charge density leads to a
strengthening of the electrostatic driving force favoring the
macrophase separation. However, note that an increase in  monomer
density leads to screening of the electrostatics, appearing through
$\tilde \kappa_{sm}$ in the expression for $D_e$. This screening
effect places an upper concentration bound on the unstable regime,
ultimately stabilizing a single isotropic phase.

\begin{figure}[ht!]
 \begin{center}
     \vspace*{1.0cm}
     \begin{minipage}[c]{15cm}
    \includegraphics[width=15cm]{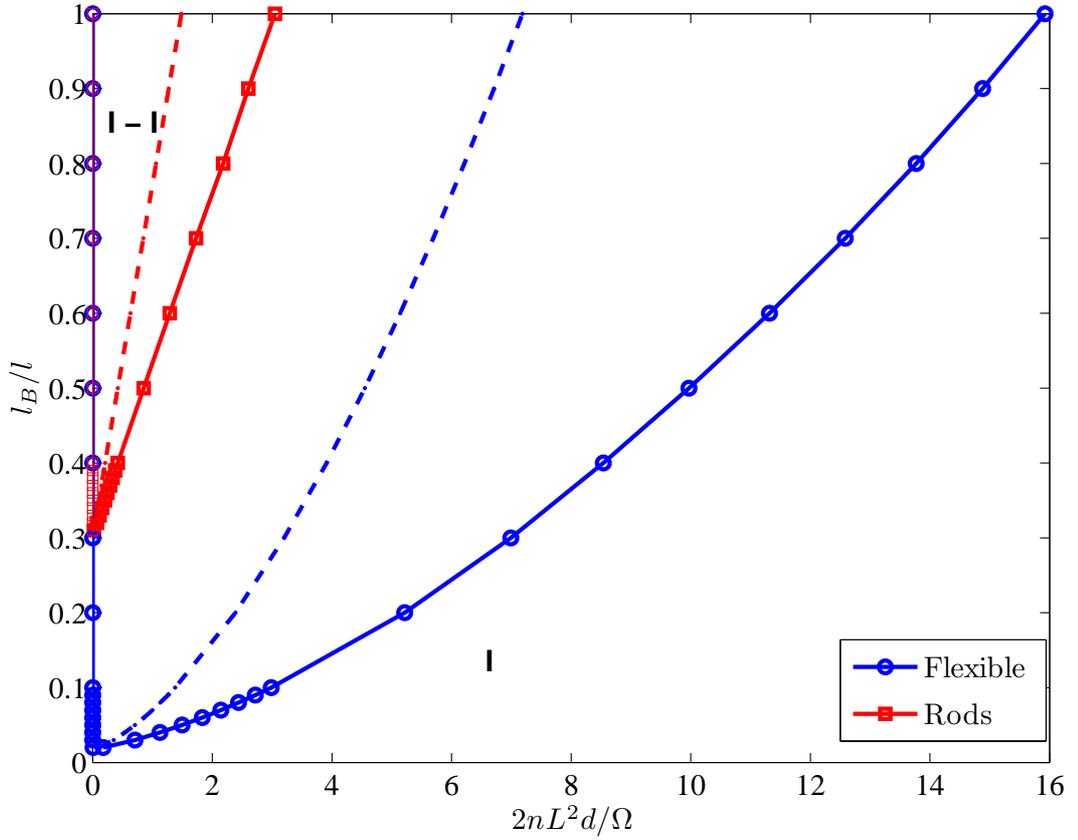}
    \end{minipage}
\caption{Comparison of the envelopes for isotropic-isotropic phase
separation in symmetric mixtures of the oppositely charged flexible
and rodlike polyelectrolytes. Dashed curves correspond to the spinodals 
and solid curves represent the binodals. The curves correspond to $N = 1000,\;
\tilde d = 1/50,\; \sigma = 0.1$. ``I'' and ``I-I'' denote one-phase
isotropic and two-phase isotropic-isotropic coexistence regimes,
respectively.} \label{fig:rods_flex_bp1}
\end{center}
\end{figure}

In Fig.~\ref{fig:rods_flex_bp1}, we have compared the spinodal phase
boundaries for flexible and rodlike symmetric polyelectrolyte
mixtures in a salt-free, counterion-free situation. It is clear that
for these weakly charged polyelectrolyte systems (characterized by
 $4\pi l_B \sigma^2 \rho < 1$) the isotropic-isotropic
coexistence regime is broader in the case of flexible symmetric
mixtures in comparison with rodlike mixtures. As all the other
contributions to the free energies of the two systems are the same
except the electrostatic contributions, it is clear that the
electrostatic driving force is stronger in the case of flexible
polyelectrolytes than rodlike polyelectrolytes. We note that these
numerical results are consistent with the simple scaling arguments
presented in the introduction.

\subsection{Full Phase Diagram}
From the results presented and discussions so far, it is clear that
electrostatics drives macrophase separation in the isotropic phase
and also stabilizes the nematic phase. So, in principle,
isotropic-isotropic, isotropic-nematic and/or nematic-nematic phase
transitions can take place in symmetric mixtures of rodlike
polyelectrolytes. Similar to the isotropic-isotropic phase
calculations considered above, we consider the extra possibilities
of nematic-isotropic and nematic-nematic coexistence, where the
orientational order parameter in each phase is determined during the
minimization of the total free energy density of the phase separated
system.

For the quasi-one component system without counterions (cf.
Eq.~\ref{eq:ftotal_iso}), we minimize the total free energy density
written as
\begin{eqnarray}
f_{total} &=& \frac{c_{p}^{\text{II}} - c_p}{c_{p}^{\text{II}} -
c_{p}^{\text{I}}}f\left\{c_{p}^{\text{I}}, \alpha^{\text{I}}
\right\} + \frac{c_p - c_{p}^{\text{I}}}{c_{p}^{\text{II}} -
c_{p}^{\text{I}}}f\left\{c_{p}^{\text{II}}, \alpha^{\text{II}}
\right\}, \label{eq:ftotal_nn}
\end{eqnarray}
where the variational parameters $\alpha^{\text{I}}$ and
$\alpha^{\text{II}}$ for the phases $\text{I}$ and $\text{II}$,
respectively, are obtained by numerically minimizing (using Brent's
method\cite{numerical_recipe}) the free energy density of the phase
($f\left\{c_{p}^{\text{I}}\right\}$ and
$f\left\{c_{p}^{\text{II}}\right\}$ for the phases $\text{I}$ and
$\text{II}$, respectively, given by Eqs.~\ref{eq:free_en_nem} -
\ref{eq:free_rods_2}) at the given monomer densities
($c_{p}^{\text{I}}$ and $c_{p}^{\text{II}}$ for the phases
$\text{I}$ and $\text{II}$, respectively). A similar approach can be
taken in quasi two-dimensional systems that include counterions. The
motivation behind carrying out such calculations is the fact that
all the phase transitions that we have considered thus far (i.e.,
isotropic-isotropic and isotropic-nematic) are subsets of these more
``general'' calculations. Unfortunately, such high dimensional
optimizations are computationally demanding, so we have looked for
opportunities to accelerate the construction of full phase diagrams.

In a complementary set of calculations, we have considered
isotropic-nematic and isotropic-isotropic phase transitions
separately. It was found that the results of the ``general''
calculations described above exactly match the results of our
isotropic-nematic and the isotropic-isotropic calculations (data not
presented here). Through such calculations, we have also established
that there are no regions of nematic-nematic coexistence, so the
various phase diagrams can be mapped out by tracking individual I-I
and I-N boundaries.

The phase diagrams presented in the remaining sections were obtained
by the above technique and are presented in
Figs.~\ref{fig:phase_diagram_rods} and \ref{fig:phase_diagram_rods_count},
respectively in the absence and presence of counterions. Overall,
the results of these calculations are in qualitative agreement with
the picture presented in Fig.~\ref{fig:phase_cartoon}. Some
important features of the full phase diagram are discussed below.

\subsubsection{Phase diagram without counterions}
\begin{figure}[ht!]
\vspace*{-0.95cm}
\begin{center}
$\begin{array}{c@{\hspace{.001in}}c@{\hspace{.001in}}}
\includegraphics[width=3.5in,height=3.5in]{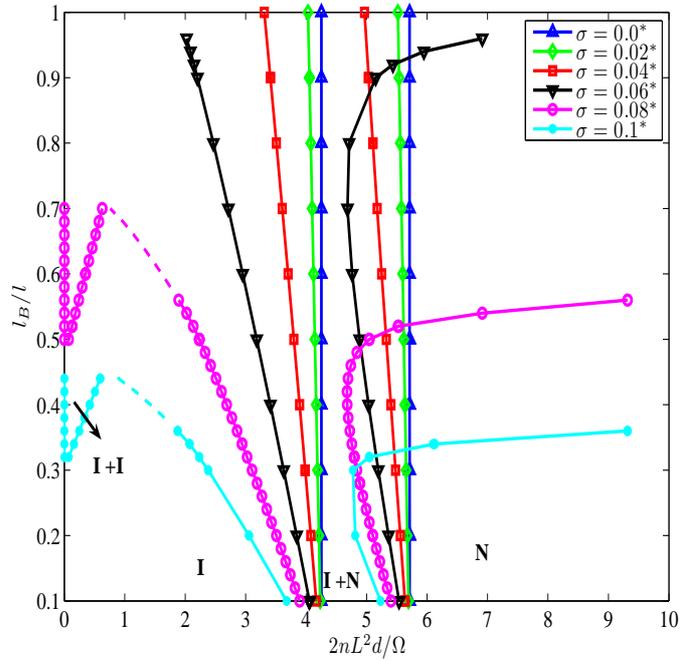}&   \\
\textbf{(a)} & \\
& \\
\includegraphics[width=3.5in,height=3.5in]{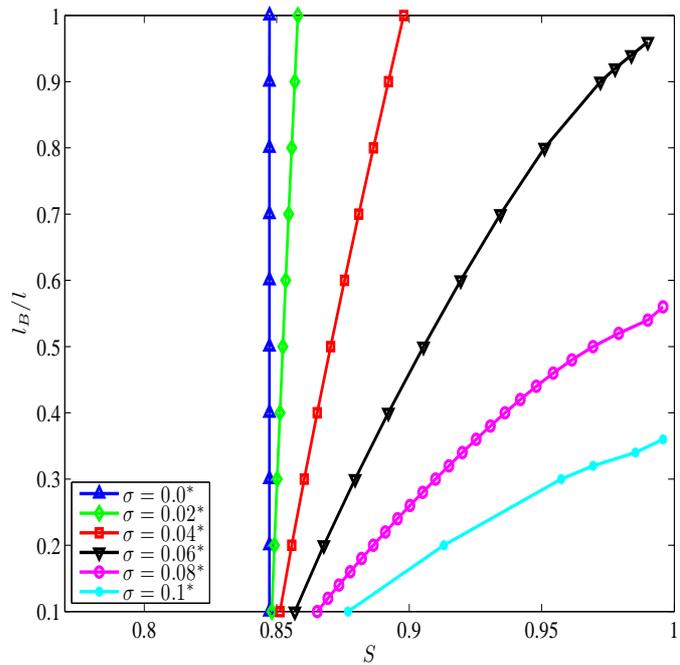} & \\
\textbf{(b)} &
\end{array}$
\vspace*{-1.0cm}
\newline
\end{center}
\caption{Phase diagram for symmetric mixtures of oppositely charged
rodlike polyelectrolytes in the absence of  counterions and for
varying linear charge density $\sigma$. Figs. (a) and (b) correspond
to the coexisting number densities of rods in the two phases and the
nematic order parameter $S$, respectively. The symbols ``I'' and
``N'' denote the isotropic and nematic phases, respectively. These
curves correspond to $N=1000,\; d/l = 1/50$. Dashed lines in Fig. (a) correspond 
to the boundary beyond which the isotropic phase becomes unstable and the completely ordered 
nematic phase becomes one of the coexisting phases.}
\label{fig:phase_diagram_rods}
\end{figure}

In Fig.~\ref{fig:phase_diagram_rods}, we have plotted the phase
diagram for symmetric mixtures of oppositely charged rodlike
polyelectrolytes in the absence of counterions. In the figure, the
linear charge density is varied to explore the effect of
electrostatics on the phase boundaries. From
Fig.~\ref{fig:phase_diagram_rods}a, it is clear that the coexisting
densities in the two phases decrease from the uncharged case with an
increase in the charge densities (compare the results for $\sigma =
0,0.02$ and $0.04$). However, a further increase in the charge
density leads to three distinct regimes corresponding to the low,
intermediate and high values of $l_B$. For very low values of the
Bjerrum length ($l_B/l \rightarrow 0$), the coexisting phases are
isotropic and nematic phases. Note that in this regime the
coexisting densities for the isotropic and nematic phases for the
polyelectrolyte systems are close to that for the neutral system.

For high values of the Bjerrum length (close to unity) and 
high charge densities such as $\sigma = 0.08, 0.1$ in 
Fig.~\ref{fig:phase_diagram_rods}a, completely ordered 
nematic phase becomes one of the coexisting phases. However, we haven't been 
able to compute the densities of the coexisting phases in this regime 
due to the numerical issues discussed in subsection F above. 

For intermediate values of $l_B$, there are regimes (e.g., $l_B/l
\in 0.1-0.6$ for $\sigma = 0.1$ in
Fig.~\ref{fig:phase_diagram_rods}a), where isotropic-isotropic
coexistence and isotropic-nematic phase separation can be separately
realized by varying the concentration of the rodlike
polyelectrolytes in solution. With an increase in $l_B$ in this
regime, three phase coexistence (isotropic-isotropic-completely ordered nematic) can be
realized in these systems. Moreover, the value of $l_B$ at which the
three phase coexistence takes place is dependent on the linear
charge density of the polyelectrolytes. In fact, the Bjerrum length
at which the three phase coexistence takes place decreases with an
increase in the linear charge density of the rods (compare the
results for $\sigma = 0.08$ and $\sigma = 0.1$ in
Fig.~\ref{fig:phase_diagram_rods}a). Note that in this intermediate
regime, the entropic effects driving the isotropic-nematic
transition are comparable in strength with the energetic effects
(coming from electrostatics), but the system quickly evolves into a
broad isotropic-nematic phase envelope upon increasing $l_B$. This
leads to a sharp increase in the density of the coexisting nematic
phase (see the plots for $\sigma = 0.08$ and $\sigma = 0.1$ in
Fig.~\ref{fig:phase_diagram_rods}a). A further increase in
$l_B$ leads to the completely ordered nematic phase 
as a coexisting phase. In order to keep track of the degree of alignment 
of the rods in the nematic phase in Fig.~\ref{fig:phase_diagram_rods}a, 
we have plotted the order parameter in Fig.~\ref{fig:phase_diagram_rods}b.

From Fig.~\ref{fig:phase_diagram_rods}b, it is clear that the order
parameter increases with an increase in $l_B/l$, which is in
agreement with the stability analysis carried out in this paper.
Hence, the numerical results support the prediction that
electrostatics favor orientational ordering. 

\subsubsection{Effect of counterions}
In contrast to the phase diagram obtained in the absence of the
counterions, the phase diagram in the presence of counterions
depends on the total number density of the rodlike polyelectrolytes
and correspondingly on the number of counterions in the system. In
order to conduct a systematic study, we have carried out two sets of
calculations. In the first set, we vary the linear charge density of
the polyelectrolytes while keeping the total number density of the
rodlike polyelectrolytes fixed at a particular value. In the second
set, the total number density of the rods is varied keeping the
linear charge density fixed at a particular value. So, in both the
sets, the total number density of counterions is varied.

\begin{figure}[ht!]
\vspace*{-0.95cm}
\begin{center}
$\begin{array}{c@{\hspace{.001in}}c@{\hspace{.001in}}}
\includegraphics[width=3.5in,height=3.5in]{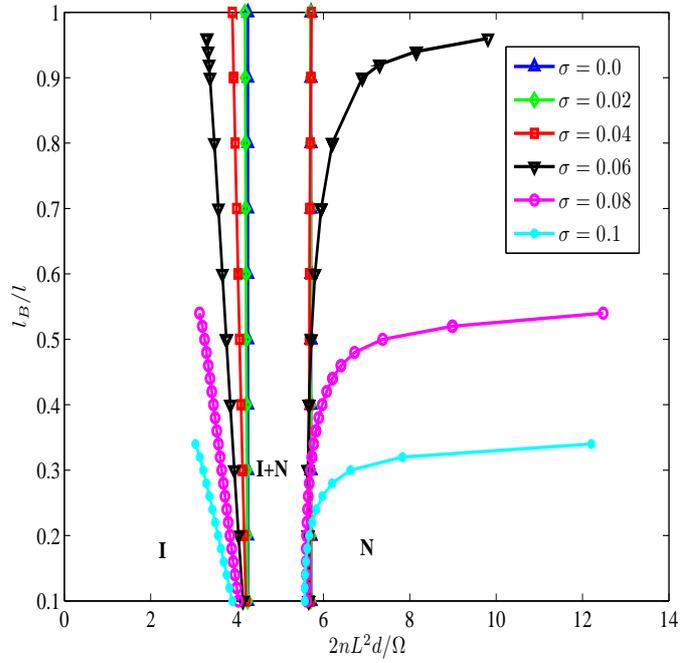}&   \\
\textbf{(a)} & \\
& \\
\includegraphics[width=3.5in,height=3.5in]{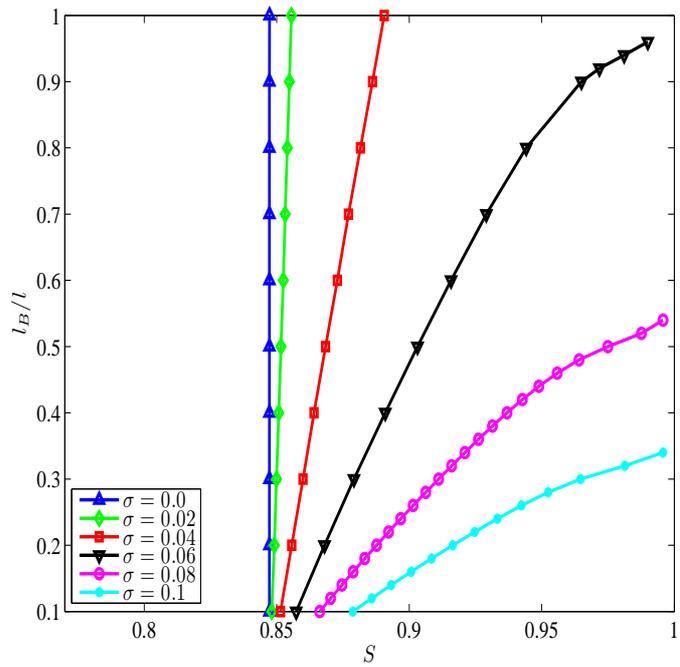} & \\
\textbf{(b)} &
\end{array}$
\vspace*{-1.0cm}
\newline
\end{center}
\caption{Phase diagram for symmetric mixtures of oppositely charged
rodlike polyelectrolytes in the presence of counterions, at fixed
total polymer density, and for varying linear charge density
$\sigma$. Figs. (a) and (b) correspond to the coexisting number
densities of the rods in the two phases and the nematic order
parameter $S$, respectively. Symbols ``I'' and ``N'' denote the
isotropic and nematic phases, respectively. These curves correspond
to $N=1000, \; d/l = 1/50, \; 2n_t L^2 d/\Omega = 4.7$.}
\label{fig:phase_diagram_rods_count}
\end{figure}

\begin{figure}[ht!]
 \begin{center}
     \vspace*{1.0cm}
     \begin{minipage}[c]{15cm}
    \includegraphics[width=15cm]{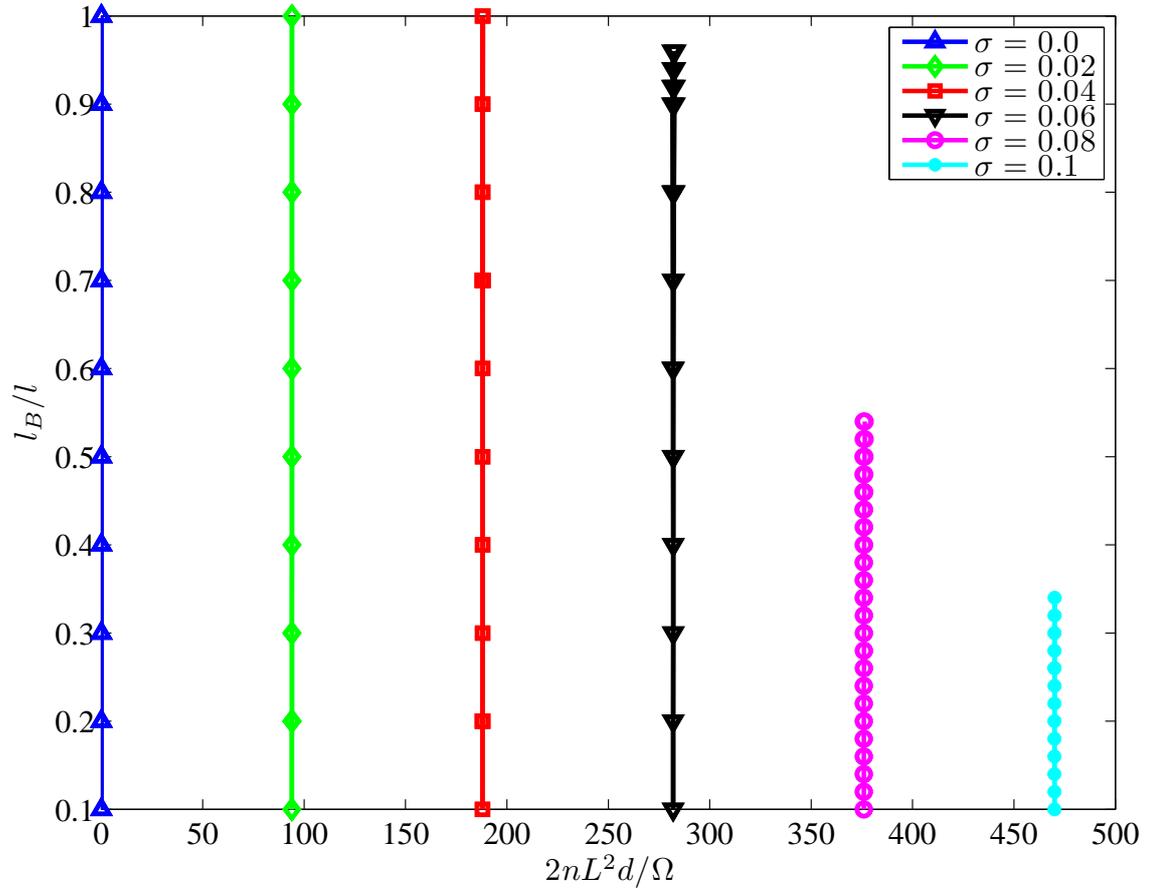}
    \end{minipage}
\caption{Counterion densities in the two coexisting phases shown in
Fig.~\ref{fig:phase_diagram_rods_count}. The counterions are uniformly distributed between the two phases 
for these set of parameters.} \label{fig:counterions_coexisting} 
\end{center}
\end{figure}

Fig.~\ref{fig:phase_diagram_rods_count} presents the results of the
first set of calculations. Comparing
Fig.~\ref{fig:phase_diagram_rods_count} with
Fig.~\ref{fig:phase_diagram_rods}, we observe that the qualitative
features of the phase diagram in the presence of counterions remain
the same as in the absence of counterions. Furthermore,
Fig.~\ref{fig:counterions_coexisting} presents the number densities
of the counterions in the coexisting phases. It is found that
counterions are uniformly distributed between the two phases
for these values of linear charge densities and the Bjerrum's length. 

\begin{figure}[ht!]
\vspace*{-0.95cm}
\begin{center}
    \begin{minipage}[c]{15cm}
    \includegraphics[width=15cm]{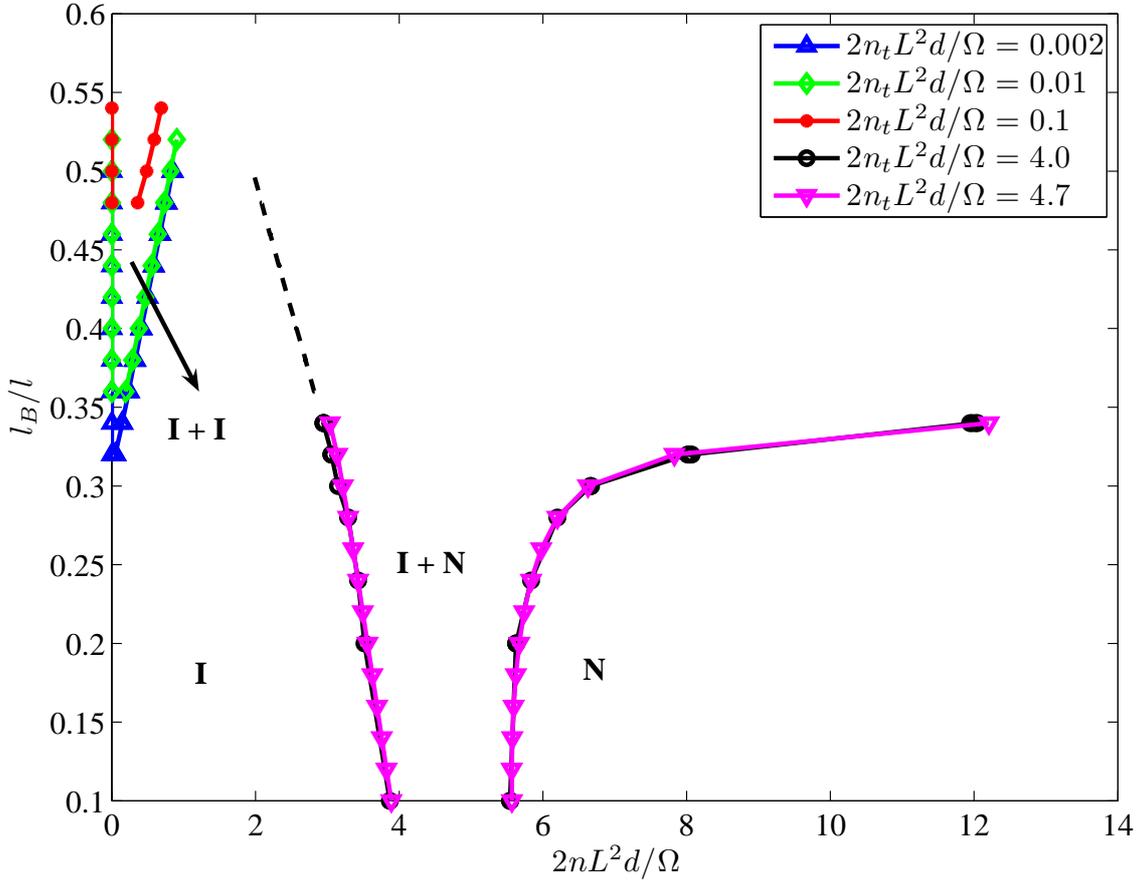}
    \end{minipage}
\vspace*{-1.0cm}
\newline
\end{center}
\caption{Effect of the total number density of rodlike
polyelectrolytes on the phase diagram in symmetric mixtures in the
presence of counterions and at fixed linear charge density. These
plots correspond to $N = 1000, \; d/l = 1/50$ and $\sigma = 0.1$. Dashed lines in Fig. (a) correspond 
to the boundary beyond which the isotropic phase becomes unstable and the completely ordered 
nematic phase becomes one of the coexisting phases.}
\label{fig:den_effect_phase}
\end{figure}

\begin{figure}[ht!]
\vspace*{-0.95cm}
\begin{center}
 \begin{minipage}[c]{15cm}
    \includegraphics[width=15cm]{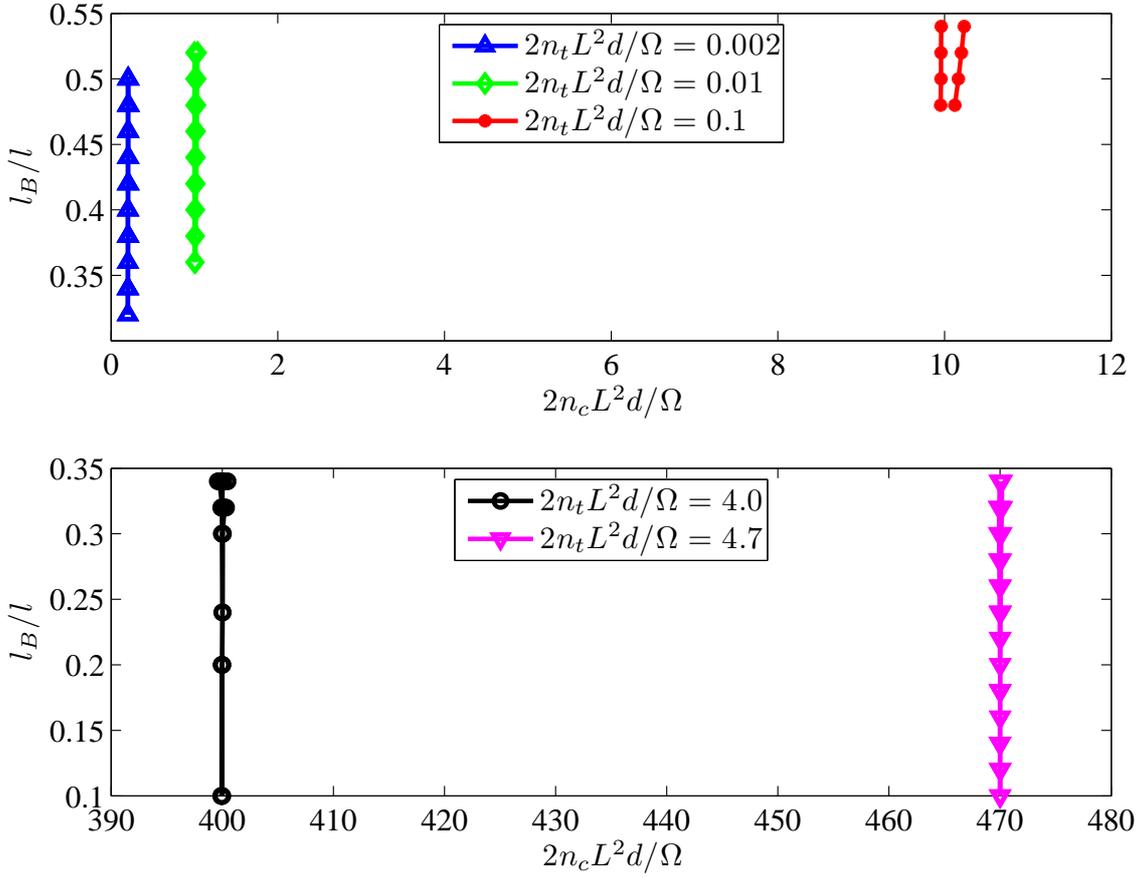}
    \end{minipage}
\vspace*{-1.0cm}
\newline
\end{center}
\caption{Counterion densities in the coexisting phases corresponding
to I-I and I-N phase transitions as shown in
Fig.~\ref{fig:den_effect_phase}. In these figures, the right branch
of each coexisting curve corresponds to the counterion density in
the phase dense in rodlike polyelectrolytes (coacervate). It is
shown that for strong enough electrostatics, the counterions also phase segregate and the counterion
density is higher in the dense phase (see the results for $2n_t L^2 d/\Omega = 0.1$). 
These plots correspond to $N = 1000, \; d/l =  1/50$ and $\sigma = 0.1$. }
\label{fig:den_effect_phase2}
\end{figure}

The results for the second set of calculations are shown in
Figs.~\ref{fig:den_effect_phase} and \ref{fig:den_effect_phase2},
where the total number density of polyelectrolytes is changed for a
particular value of the linear charge density. These results show
the isotropic-isotropic and isotropic-nematic coexistence regimes
just like the ones seen in the absence of counterions in
Fig.~\ref{fig:phase_diagram_rods}. In these calculations, the change
in the total number densities of the polyelectrolytes leads to a
change in the total number density of the counterions. The effect of
varying the total number density of the polyelectrolytes and
counterions on the phase boundaries can be explained as follows. As
already discussed, the origin of the isotropic-nematic transition
for neutral rods near the overlap concentration or for very low
values of $l_B$ ($l_B/l \rightarrow 0$) is entropic and is nearly
independent of electrostatics. Hence, the presence of counterions
does not affect these boundaries (compare the phase boundaries for
$2n_t L^2 d/\Omega = 4$ and $4.7$). However, the origin of the
isotropic-isotropic transition 
at low number
density of polyelectrolytes is electrostatically driven and is
strongly dependent on the presence of the counterions. The width of
this particular coexistence regime can be tuned by changing the
number of counterions. In particular, the coexistence regimes shrink
with an increase in the number of counterions due to the screening
of electrostatic interactions(compare $2n_t L^2 d/\Omega = 0.002,
0.01$ and $0.1$ in Fig.~\ref{fig:den_effect_phase}). Note that the
results for the isotropic-isotropic phase transition in
Fig.~\ref{fig:den_effect_phase} are the same as in
Fig.~\ref{fig:ii_bet_n1000}.

Figure~\ref{fig:den_effect_phase2} shows the counterion distribution
in the coexisting phases. Consistent with our prior results, it is
evident that the counterions also phase segregate at high Bjerrum
length with a higher density in the concentrated (coacervate) phase.

\section{Conclusions}\label{sec:conclusions}

We have studied the phase behavior of salt-free symmetric mixtures
of oppositely charged rodlike polyelectrolytes using the random
phase approximation. In this work, we have focused on weak 
polyelectrolytes in the regime $l_B/l < 1$ to avoid complications
arising from possible counterion condensation. For a variety of
symmetric mixtures, we have computed the phase boundaries for
regions of isotropic-isotropic and isotropic-nematic coexistence. We
were not able to identify any regions of nematic-nematic coexistence
in these symmetric systems.

Our stability analysis and numerical results for coexistence curves
reveal that electrostatic interactions favor nematic ordering of
the rodlike components in solution.  Nonetheless, the screening of
these electrostatic interactions by higher concentrations of
counterions weakens or destroys this ordering. It is shown that the
phase boundaries for symmetric mixtures containing oppositely
charged rodlike polyelectrolytes are dependent on the electrostatic
interaction strengths characterized by the linear charge density,
$\sigma$, and the Bjerrum length, $l_B$. In particular, it is
demonstrated that at low electrostatic interaction strengths (i.e.,
$l_B/l \rightarrow 0, \: \sigma \rightarrow 0$), the densities in
the coexisting isotropic and nematic phases decrease with an
increase in the linear charge density of the polyelectrolytes in the
absence of counterions. However, an increase in the electrostatic
interaction strength by increasing $\sigma$ leads to three distinct
regimes characterized by $l_B/l$. At low $l_B /l$ (close to zero), a
narrow region of isotropic-nematic coexistence prevails, whose
origin lies in the entropy of the system. On the other hand, at
relatively high $l_B /l$ (close to unity), the completely ordered nematic phase 
becomes one of the coexisting phases. 
However, its origin lies in the electrostatic attraction between
oppositely charged polyelectrolytes. At intermediate values of
$l_B$, isotropic-isotropic or isotropic-nematic coexistence can
prevail depending on the concentration of the polyelectrolytes.
Also, in this regime, at a particular value of $l_B$, three phases
(low density isotropic, moderate density isotropic, and high density
nematic) coexist with each other. The value of $l_B$ at which the
three phase coexistence takes place depends sensitively on the
linear charge density of the polyelectrolytes.

We have also investigated the effect of counterions on the phase
coexistence boundaries (isotropic-isotropic and isotropic-nematic).
Comparison of the results for the systems with and without
counterions reveals that sufficiently high concentrations of
counterions suppress both isotropic-isotropic and isotropic-nematic
coexistence. A key prediction of theory is the result that the
concentration of counterions in the dense (or coacervate) phase is
slightly higher than in the dilute (or supernatant) phase. Also,
comparison of the phase separation boundaries between comparable
rodlike and  flexible polyelectrolyte mixtures reveals that the
isotropic-isotropic macrophase separation regime is broader in the
case of weakly charged flexible polyelectrolytes.

Furthermore, in this work, we have limited ourselves to the phase separation regime 
in the mixtures containing weakly charged polyelectrolytes close to the critical point. We have found 
that the critical points for the 
isotropic-isotropic and isotropic-nematic phase transitions exist at 
very low number density of rods ($n/\Omega \sim 1/L^2 d$) and weak electrostatic interaction 
strengths (i.e., $l_B/l \ll 1$). At this point, we remark on the range of validity of the 
theory to describe the phase boundaries and some of the future directions. There are three key issues, which limits the validity of the theory. In future, we'll extend the theory to higher electrostatic interaction strengths by 
addressing the issues mentioned below. 

First issue is the use of the random phase approximation (RPA) to compute the contribution to the free energy coming from the electrostatic correlations. A typical way to judge the validity of the RPA is to compare the mean field contribution with the correlation term in the free energy. Due to the fact that the mean field contribution to the 
free energy coming from the electrostatics is zero and correlation terms beyond the RPA are not available, 
the range of validity of RPA can't be inferred directly. However, general consensus\cite{borue_complexation,muthu_double,joanny_leibler,kaji_kanaya} is that the 
RPA for flexible polymers is valid for concentrations above the overlap concentration. In this work, the coacervate 
phase has concentration above the overlap concentration and the supernatant phase is very dilute. Indeed, the 
supernatant phase is not well described by the RPA. On the other hand, the RPA is suitable for a very good 
quantitative description of the phase boundary describing the coacervate phase. Remarkably, 
the RPA predicts a supernatant phase with almost zero density in qualitative agreement with the 
experiments\cite{dubin_review,pe_complex_reviews,voorn_review,veis_complexation,kabanov_work1,kabanov_work2,dautzenberg_work1,pogodina_work,dautzenberg_work2}. Furthermore, the RPA describing the coacervate phase boundary in the case of flexible polyelectrolyte mixtures has been compared extensively with the experiments\cite{leibler_borukhov} and simulations\cite{jay_yuri_2}. Indeed, agreement between the theory, 
simulations and experiments is remarkable. In order to go beyond the RPA, Hartree approximation\cite{glenn_book} can be used, which can provide a more quantiative description for the supernatant phase also. However, we leave it for future work.

Second issue arises due to the use of the Onsager second virial approach to describe the steric effects in the case of 
long rods. It is well-known\cite{edwardsbook,straley_review} that the approach only works in the limit of very long aspect ratio of the rods. In fact, the limit of validity of the Onsager corresponds to $L/d \gg 10$. Furthermore, the second virial approach is strictly valid for low number density of the rods. For dense systems, higher order terms needs to be considered. For very long rods, the isotropic-nematic and isotropic-isotropic phase transitions occur at low enough number densities of the rods (which is of the order of overlap concentration). This is the reason 
the second virial approach is able to correctly predict these phase transitions at low electrostatic interaction strengths. However, an increase in the electrostatic interaction strength causes the density of the coacervate phase to increase and the virial approach breaks down. This deficiency of the theory can be removed by considering the effect of higher order terms\cite{parsons_paper,lee_higher}, which we'll consider in future. 

Third issue is the ignorance of charge renormalization while describing phase separation. In this work, the  
phase separation regime corresponding to very low electrostatic interaction strengths is described assuming that there is no counterion adsorption on the polyelectrolytes. In this regime, the issue of charge renormalization during phase separation can be safely ignored. However, with the increase in the electrostatic interaction strength, counterions can adsorb on the polyelectrolytes and modulate their charge. For this part of the phase diagram, the degree of ionization for each polyelectrolyte in each phase needs to be computed, while equating the chemical potential and pressure of each component in the two phases\cite{levin_charge_regularization,muthu_charge_regularization}. In other words, the possibility of different degree of ionizations in daughter phases needs to be explored. In this work, we have limited the theoretical study to $l_B/l < 1$, where issue of counterion adsorption can be safely ignored. 

A related issue is the consideration of ionic clusters (such as dipoles, quadrapoles etc.) formed as a result of strong electrostatic attraction between different oppositely charged components. It has been shown\cite{fisher_levin,fisher} in the literature that the issue of ion-pairing has to be taken into account in the case of 
hard sphere model for simple electrolytes (also known as restricted primitive model (RPM)) 
to correctly match the simulation results in the critical regime. 
This is a manifestation of the fact that the critical point for the RPM, as predicted by the Debye-H\"{u}ckel 
theory, corresponds\cite{fisher_levin} to $q^2 l_B^\star/a = 16$, where $q$ is the charge per sphere, $l_B^\star$ is the Bjerrum length 
at the critical point and $a$ is the diameter of the hard spheres. For symmetrical electrolytes, $a$ is the same for both kinds of charged spheres. Also, for monovalent electrolytes, $q=1$ and the critical point exists at very low temperature such that $l_B^\star/a = 16$. At such low temperatures, indeed one has to extend the Debye-H\"{u}ckel 
theory (which is a RPA like) by including atleast dipolar interactions in the regime near 
the critical point. However, in contrast to the symmetric electrolytic mixtures, the phase separation regime close to the critical point in the case of mixtures containing long flexible or rodlike polyelectrolytes can be well described within RPA without any consideration of ion-ion, ion-polyelectrolyte (which is the same as the counterion adsorption) or polyelectrolyte-polyelectrolyte pairs. This is an outcome of the fact that now the phase separation takes place at relatively higher temperatures and low densities. In this regime, the mixtures of oppositely 
charged polyelectrolytes behave as weakly correlated liquids\cite{rubinstein_pairing}. However, if we increase the electrostatic interaction strength and go far from the critical regime, there is a competition for pairing between different oppositely charged species. A rough estimate for the minimum driving force for pairing between the ions is given by Bjerrum's theory\cite{ebeling_work, levin_review} of ion-pairs. According to the theory, life time of the 
paired state in the case of two oppositely charged ions (while undergoing thermal motion) significantly increases when $q^2 l_B/r > 2$, $r$ being the distance between the ions. In this regime, the electrostatic 
attraction takes over the thermal energy of ions and ion-ion pair (or dipole) needs to be considered as a new species. 
This relation can be cast in terms of number density of ions ($= n_c/\Omega$) by using $r \sim (n_c/\Omega)^{-1/3}$ so that $q^2l_B/l > 1/(n_c/\Omega)^{1/3}$. Issue of ion pairs/clusters formed as a result of binding, which involves 
charged monomers is more suble compared to the issue of pairing in small ions. 

However, we can estimate the regime, where one has to explicitly consider the binding between 
the polyelectrolytes and oppositely charged counterions or polyelectrolytes. Carrying out a single chain analysis, it has been shown that the binding of counterions on flexible and rodlike polyelectrolytes becomes important\cite{manning,counterion_adsorption_muthu} roughly around $l_B/l > 1$. Similarly, an analysis of the system containing two oppositely charged flexible polyelectrolytes\cite{zhaoyang_complex} reveals that the complexation takes place only for strong electrostatic interaction strengths (i.e., $\sigma^2 l_B > 1$). For rodlike polyelectrolytes, the electrostatic interaction strength required for complexation needs to be stronger compared to the flexible polyelectrolytes due to weaker (logarithmic) electrostatic potential for rodlike polymers. These single chain analyses provide a clear picture about the dilute solution regime. In the regime above the overlap concentration for polymers, situation is 
more complicated and one has to consider the multi-chain effects. However, one can carry out a mean-field analysis\cite{monica_complexation_2,semenov_rubinstein} to estimate the fraction of charged monomers involved in binding (say, $\Gamma$). It can be readily shown\cite{monica_complexation_2} that the fraction is given by the relation $\Gamma/(1-\Gamma)^2 = (\sigma/l )\left[c_p/(2 Ld)\right]\exp(|E/k_B T|)$, $E/k_BT \sim l_B$ being the energy gain \textit{per pair}. In this work, we have considered weakly charged polyelectrolyte solutions at very low electrostatic interaction strengths so that the fraction of charged monomers involved in pairing is close to zero.

At present we are not aware of experimental data sets sufficient for
a comprehensive test of the theoretical predictions made here. With
this aim, we welcome interactions with experimental groups to define
appropriate systems and experimental protocols.

\section*{ACKNOWLEDGEMENT}
\setcounter {equation} {0} \label{acknowledgement} We are grateful
to Prof. Philip A. Pincus and Dr. Yongseok Jho for useful discussions on the phase
behavior of polyelectrolytes. Financial support was provided
by the UCSB-MIT-Caltech Institute for Collaborative Biotechnologies 
and the Materials Research Laboratory (MRL) at UCSB..
This work made use of the MRL Computing Facilities supported by the
MRSEC Program of the National Science Foundation under Award No.
DMR05-20415.

\clearpage
\renewcommand{\theequation}{A-\arabic{equation}}
  \setcounter{equation}{0}  
 \section*{APPENDIX A : Partition function for mixtures of rodlike polyelectrolytes } \label{app:A}

Here, we present the details of our derivation of the free energy for the mixtures of rodlike
polyelectrolytes in the presence of counterions as described in Section II. Different parameters representing 
the number of rods and counterions, length and diameter of the rods, and charge along the rods have already 
been described in Section II. In terms of these parameters, 
the partition function can be written as
\begin{eqnarray}
Z &=&  \frac{\prod_{k=+,-}\prod_{j=1}^{n_k}\left[\int_{p_k(\mathbf{u}_{j},\nh)} d\mathbf{r}_{j}d\mathbf{u}_{j} \right]}{\prod_{k=+,-}n_k!\prod_{j' =  c+,c-}n_{j'}!} \prod_{j' =  c+,c-} \prod_{j'' =  1}^{n_{j'}} \int  d\mathbf{r}_{j''}\exp \left [- \frac{H_{int}}{k_BT} \right ]
\end{eqnarray}
where 
\begin{eqnarray}
       \frac{H_{int}}{k_BT} &=&  \frac{H_{w}}{k_BT} + \frac{H_{e}}{k_BT}, \\
      \frac{H_{w}}{k_BT} &=& \frac{1}{2}\sum_{j=1}^{n_+ + n_-}\sum_{k=1}^{n_+ + n_-} W(\mathbf{r}_j,\mathbf{u}_{j},\mathbf{r}_k,\mathbf{u}_{k}), \\
\frac{H_{e}}{k_BT} &=& \frac{l_B}{2}\int d\mathbf{r} \int d\mathbf{r}' \frac{\hat{\rho}_e(\mathbf{r})\hat{\rho}_e(\mathbf{r}')}{|\mathbf{r} - \mathbf{r}'|}.
\end{eqnarray}
Here, $p_k(\mathbf{u}_{j},\nh)$ for $k=+,-$ is the probability of finding a rod of type $k$ oriented along 
unit vector $\mathbf{u}_j$ when the director is taken to be along the unit vector $\nh$. Subscript $p_k(\mathbf{u},\nh)$ 
under the integral symbol means that the integration need to be carried out under the 
constraint that the orientational distribution function for the positive and negative rods are $p_+(\mathbf{u},\nh)$ 
and $p_-(\mathbf{u},\nh)$, respectively. Also,  
hamiltonian $H_{int}$ is divided into contributions coming from 
the short range excluded volume interactions ($H_w$) and the long range electrostatic interactions ($H_e$). 
Furthermore, in the expression for $H_w, W(\mathbf{r}_j,\mathbf{u}_{j},\mathbf{r}_k,\mathbf{u}_{k})$ 
characterizes the excluded volume interactions between two rods whose centers are located at 
$\mathbf{r}_j$ and $\mathbf{r}_k$ with their axes oriented along $\mathbf{u}_j$ and $\mathbf{u}_k$, respectively. Also, 
$l_B$ is Bjerrum length and $\hat{\rho}_e$ in the expression for $H_e$ is the local charge density defined as
        \begin{eqnarray}
      \hat{\rho}_e(\mathbf{r}) &=& Z_{+}\sigma_+ \hat{\rho}_+(\mathbf{r}) + Z_{-}\sigma_- \hat{\rho}_-(\mathbf{r}) + \sum_{j' = c+,c-} Z_{j'} \hat{\rho}_{j'}(\mathbf{r}) \\
    \hat{\rho}_{k}(\mathbf{r})  &=& N_k \sum_{j=1}^{n_{k}}\frac{1}{L_k}\int_{-L_k/2}^{L_k/2} ds_{j} \, \delta (\mathbf{r}_{j} + s_{j} \mathbf{u}_{j} - \mathbf{r}), \quad \mbox{for} \quad k = +,-,\\
     \hat{\rho}_{j'}(\mathbf{r}) &=&  \sum_{j''=1}^{n_{j'}} \delta (\mathbf{r}-\mathbf{r}_{j''}), \quad \mbox{for} \quad j' = c+,c-,
\end{eqnarray}
where $Z_{\gamma}$ is the valency (with sign) of the charged species of type $\gamma$ and $N_k$ is the 
number of monomers for the rod of type $k$ so that $L_k = N_k l_k$, $l_k$ being the monomer length. Also, $s_j$ 
in the expression for $\hat{\rho}_k(\mathbf{r})$ is the contour variable used to locate any monomer along the $j^{th}$ 
rod of type $k$. In order to proceed further, we rewrite $Z$ as 
\begin{eqnarray}
       Z& = & Z_o \left < \exp \left [- \frac{H_{int}}{k_BT} \right ]\right >_{p_+,p_-} \label{eq:parti}
\end{eqnarray}
where $Z_o$ is the partition function for a non-interacting system with the same orientational distributions 
(i.e., $p_+,p_-$) and also the normalization factor in the expression for $Z$, given by 
\begin{eqnarray}
      Z_o &=& \exp\left[- \frac{F_o}{k_BT}\right]  = \frac{\prod_{k=+,-}\prod_{j=1}^{n_k}\left[\int_{p_k(\mathbf{u},\nh)}d\mathbf{r}_{j}d\mathbf{u}_{j} \right]}{\prod_{k=+,-}n_k!\prod_{j' =  c+,c-}n_{j'}!}\int \prod_{j' =  c+,c-} \prod_{j'' =  1}^{n_{j'}} d\mathbf{r}_{j''} \nonumber \\
      &=& \prod_{\gamma = +,-,c+,c-}\left[\frac{\Omega^{n_\gamma}}{n_\gamma!}\right]\prod_{k=+,-}\prod_{j=1}^{n_k} \int_{p_k(\mathbf{u},\nh)}d\mathbf{u}_{j} . 
      \end{eqnarray}
Using Stirling's approximation for factorials $ n! \simeq n (\ln n -  1)$ and considering different ways of distributing $n_k$ rods for a given orientational probability distribution function $p_k(\mathbf{u}_{j},\nh)$ 
along the surface of a unit sphere\cite{edwardsbook}
\begin{eqnarray}
\frac{F_o}{k_BT} & = & \sum_{j=+,-}n_j \int
d\mathbf{u}_{j} p_j(\mathbf{u}_{j},\nh) \ln \left[4\pi
p_j(\mathbf{u}_{j},\nh)\right] + \sum_{\gamma = +,-,c+,c-} n_{\gamma} \left[\ln \frac{n_{\gamma}}{\Omega} - 1\right].
\end{eqnarray}
Also,
\begin{eqnarray}
    \left<\exp \left [- \frac{H_{int}}{k_BT} \right ]\right>_{p_+,p_-} &=& \frac{\prod_{k=+,-}\prod_{j=1}^{n_k}\left[\int_{p_k(\mathbf{u}_{j},\nh)} d\mathbf{r}_{j}d\mathbf{u}_{j} \right]\prod_{j' =  c+,c-} \prod_{j'' =  1}^{n_{j'}} \int d\mathbf{r}_{j''}\exp \left [- \frac{H_{int}}{k_BT} \right ] }{\prod_{k=+,-}\prod_{j=1}^{n_k}\left[ \int_{p_k(\mathbf{u}_{j},\nh)} d\mathbf{r}_{j}d\mathbf{u}_{j}\right]\prod_{j' =  c+,c-} \prod_{j'' =  1}^{n_{j'}} \int d\mathbf{r}_{j''}}. \nonumber \\
&&
\end{eqnarray}
Using the Hubbard-Stratonovich transformation\cite{glenn_book} for the electrostatic part i.e., $H_e$ and using 
Fourier transform defined for any arbitrary function $f(\mathbf{r})$ by  
\begin{eqnarray}
       f(\mathbf{r}) &=& \int \frac{d^{3}\mathbf{q}}{(2\pi)^3}f_q e^{i \mathbf{q}\cdot \mathbf{r}},
\end{eqnarray}
the partition function can be written as
\begin{eqnarray}
       \frac{Z}{Z_o} &=& \frac{ \int \prod_{q}D[\psi_q] \exp\left[-\frac{1}{2}\int \frac{d^{3}\mathbf{q}}{(2\pi)^3}\psi_q \frac{q^2}{4\pi l_B}\psi_{-q} + \sum_{j' = c+,c-}n_{j'}\ln Q_{j'}\left\{\psi_q\right\} \right ] }{\int \prod_{q}D[\psi_q]\exp\left[-\frac{1}{2}\int \frac{d^{3}\mathbf{q}}{(2\pi)^3}\psi_q \frac{q^2}{4\pi l_B}\psi_{-q}\right ]  } \nonumber \\
&& \frac{\prod_{k=+,-}\prod_{j=1}^{n_k}\left[\int_{p_k(\mathbf{u}_{j},\nh)}d\mathbf{r}_{j}d\mathbf{u}_{j} \right] \exp \left [- \frac{H_{w}}{k_BT} + i \sum_{k=+,-}\int \frac{d^{3}\mathbf{q}}{(2\pi)^3}\psi_q Z_k \sigma_k \rho_{k,-q}\right] }{\prod_{k=+,-}\prod_{j=1}^{n_k}\left[\int_{p_k(\mathbf{u}_{j},\nh)} d\mathbf{r}_{j}d\mathbf{u}_{j}\right] }\nonumber \\
&& \label{eq:parti_all}
\end{eqnarray}
where 
  \begin{eqnarray}
\rho_{k,q}  &=& \int\, d\mathbf{r}  \hat{\rho}_k(\mathbf{r}) e^{i\mathbf{q}\cdot\mathbf{r}} 
= \sum_{i=1}^{n_{k}}N_k e^{i\mathbf{q}.\mathbf{r}_i}  \frac{\sin\left[\left(\mathbf{q}\cdot\mathbf{u}_i\right)L_k/2\right]}{\left[\left(\mathbf{q}\cdot\mathbf{u}_i\right)L_k/2\right]}, \quad \mbox{for} \quad k = +,- \label{eq:fourier_den}
\end{eqnarray}
and
\begin{eqnarray}
       Q_{j'}\left\{\psi_q \right\} & = & \frac{1}{\Omega}\int d\mathbf{r} \exp \left[i \int \frac{d^{3}q}{(2\pi)^3}\psi_q Z_{j'}e^{i\mathbf{q}\cdot\mathbf{r}} \right] \label{eq:partition_smallions}
\end{eqnarray}
So far the partition function is exact. However, evaluation of the exact partition function is a tedious task. Useful 
insights can be obtained by invoking the following approximations. 

From Eq. ~\ref{eq:partition_smallions}
\begin{eqnarray}
\sum_{j' = c+,c-}\ln Q_{j'}^{n_{j'}}\left\{\psi_q\right\} &\simeq&  \sum_{j' = c+,c-}\ln \left[1 - \frac{Z_{j'}^2 n_{j'}}{2\Omega}\int \frac{d^{3}q}{(2\pi)^3}\psi_q \psi_{-q}\right ] \nonumber \\
&\simeq &  - \sum_{j' = c+,c-}\frac{Z_{j'}^2 n_{j'}}{2\Omega}\int \frac{d^{3}q}{(2\pi)^3}\psi_q \psi_{-q}. \label{eq:log_exp}
\end{eqnarray}
Note here that the linear terms arising from the expansion of the exponential in $Q_{j'}$ vanish in the sum due to 
the presence of oppositely charged counterions, i.e. $Z_{c+} = - Z_{c-}$. Furthermore, approximating 
the logarithm by the first term in the expansion in Eq. ~\ref{eq:log_exp} is strictly valid for dilute 
concentrations of the counterions so that $n_{j'}/\Omega$ is small. Using this approximation, 
Eq. ~\ref{eq:parti_all} becomes
\begin{eqnarray}
       \frac{Z}{Z_o} &=& \frac{ \int \prod_{q}D[\psi_q] \exp\left[-\frac{1}{8\pi l_B}\int \frac{d^{3}\mathbf{q}}{(2\pi)^3}\psi_q \left(q^2 + \kappa^2\right)\psi_{-q} \right ] }{\int \prod_{q}D[\psi_q]\exp\left[-\frac{1}{8\pi l_B}\int \frac{d^{3}\mathbf{q}}{(2\pi)^3}\psi_q q^2\psi_{-q}\right ]  } \nonumber \\
&& \frac{\prod_{k=+,-}\prod_{j=1}^{n_k}\left[\int_{p_k(\mathbf{u}_{j},\nh)}d\mathbf{r}_{j}d\mathbf{u}_{j} \right] \exp \left [- \frac{H_{w}}{k_BT} + i \sum_{k=+,-}\int \frac{d^{3}\mathbf{q}}{(2\pi)^3}\psi_q Z_k \sigma_k \rho_{k,-q}\right] }{\prod_{k=+,-}\prod_{j=1}^{n_k}\left[\int_{p_k(\mathbf{u}_{j},\nh)} d\mathbf{r}_{j}d\mathbf{u}_{j}\right] }\nonumber \\
&& \label{eq:parti_all2}
\end{eqnarray}
where we have defined $\kappa^2 = 4\pi l_B \sum_{j'} Z_{j'}^2 n_{j'}/\Omega $ so that $\kappa^{-1}$ is the Debye length. Now the integrals over the spatial and orientational degrees of freedom under the constraint of the 
given distribution functions can be carried out using the approximation described in Ref. \cite{edwardsbook}. 
To be explicit, 
\begin{eqnarray}
       I_{p}\left\{\psi_q\right\} &=& \frac{\prod_{k=+,-}\prod_{j=1}^{n_k}\left[\int_{p_k(\mathbf{u}_{j},\nh)}d\mathbf{r}_{j}d\mathbf{u}_{j} \right] \exp \left [- \frac{H_{w}}{k_BT} + i \sum_{k=+,-}\int \frac{d^{3}\mathbf{q}}{(2\pi)^3}\psi_q Z_k \sigma_k \rho_{k,-q}\right] }{\prod_{k=+,-}\prod_{j=1}^{n_k}\left[\int_{p_k(\mathbf{u}_{j},\nh)} d\mathbf{r}_{j}d\mathbf{u}_{j}\right] }\nonumber \\
&=& \frac{\prod_{k=+,-}\prod_{j=1}^{n_k}\left[\int_{p_k(\mathbf{u}_{j},\nh)}d\mathbf{r}_{j}d\mathbf{u}_{j} \right] \left[1 - \left\{1 - \exp \left (- \frac{H_{w}}{k_BT}\right) \right\}\right] \exp\left[i \sum_k\int \frac{d^{3}\mathbf{q}}{(2\pi)^3}\psi_q Z_k \sigma_k \rho_{k,-q}\right] }{\prod_{k=+,-}\prod_{j=1}^{n_k}\left[\int_{p_k(\mathbf{u}_{j},\nh)} d\mathbf{r}_{j}d\mathbf{u}_{j}\right] } \nonumber \\
&=& 1 -  \frac{\prod_{k=+,-}\prod_{j=1}^{n_k}\left[\int_{p_k(\mathbf{u}_{j},\nh)}d\mathbf{r}_{j}d\mathbf{u}_{j} \right] \left\{1 - \exp \left (- \frac{H_{w}}{k_BT}\right) \right\}}{\prod_{k=+,-}\prod_{j=1}^{n_k}\left[\int_{p_k(\mathbf{u}_{j},\nh)} d\mathbf{r}_{j}d\mathbf{u}_{j}\right]} \nonumber \\
&& - \frac{1}{2}\frac{\prod_{k=+,-}\prod_{j=1}^{n_k}\left[\int_{p_k(\mathbf{u}_{j},\nh)}d\mathbf{r}_{j}d\mathbf{u}_{j} \right]\left[\int \frac{d^{3}\mathbf{q}}{(2\pi)^3} \psi_q \sum_k Z_k \sigma_k \rho_{k,-q}\right]^2 } {\prod_{k=+,-}\prod_{j=1}^{n_k}\left[\int_{p_k(\mathbf{u}_{j},\nh)} d\mathbf{r}_{j}d\mathbf{u}_{j}\right]} + \cdots \label{eq:orientation_int}
\end{eqnarray}
where we have ignored a cross-term between the excluded volume interactions and the electrostatics part to keep the calculations analytically tractable. Now, assuming that the excluded volume interactions between the rods occur 
independently of each other, second term in the series in Eq. ~\ref{eq:orientation_int} can be 
evaluated.\cite{edwardsbook} Third term in the series is also straigthforward to evaluate after plugging 
the expression for $\rho_{k,q}$ given in Eq. ~\ref{eq:fourier_den}. Now, exponentiating the series

\begin{eqnarray}
       \frac{Z}{Z_o} &=&  \exp\left[-\frac{1}{2\Omega} \sum_{j,k=+,-}' n_j n_k \int d\mathbf{u} \int d\mathbf{u}' p_j(\mathbf{u},\nh)
\left [2L_j L_k d |\mathbf{u}\times\mathbf{u}'| \right]p_k(\mathbf{u}',\nh) \right]\nonumber \\
&&  \frac{ \int \prod_{q}D[\psi_q] \exp\left[-\frac{1}{8\pi l_B}\int \frac{d^{3}\mathbf{q}}{(2\pi)^3}\psi_q \left\{q^2 + \kappa^2
+ \sum_{k=+,-}\kappa_k^2 t_k(qL_k)\right\}\psi_{-q}\right ] }{\int \prod_{q}D[\psi_q]\exp\left[-\frac{1}{8\pi l_B}\int \frac{d^{3}\mathbf{q}}{(2\pi)^3}\psi_q q^2 \psi_{-q}\right ] } \nonumber \\
&&
\end{eqnarray}
where we have defined $\kappa_k^2 = 4\pi l_B Z_{k}^2 \sigma_{k}^2 L_k^2 n_k/\Omega$ and 
 \begin{eqnarray}
    t_k(q L_k)  &=& \int d\mathbf{u} p_{k}(\mathbf{u},\nh)\left[\frac{\sin\left[\left(\mathbf{q}.\mathbf{u}\right)L_k/2\right]}{\left[\left(\mathbf{q}.\mathbf{u}\right)L_k/2\right]}\right]^2, \quad \mbox{for} \quad k = +,-.
\end{eqnarray}
Also, the primed superscript means that the $j=k$ terms are omitted from the double sum. 
By carrying out the Gaussian functional integrals over $\psi_q$ and
subtracting out the free energy in the low density limit,
Eq.~\ref{eq:free_rods} is obtained.

\renewcommand{\theequation}{B-\arabic{equation}}
  \setcounter{equation}{0}  
 \section*{APPENDIX B : Limiting cases for the electrostatic part of the free energy of the isotropic phase} \label{app:B}

Here, we provide a derivation of the approximate form of the
electrostatic contribution to the free energy of the isotropic phase
in the limiting cases of $L_k \rightarrow \infty$ (cf.
Eqs.~\ref{eq:fe_limit1} and ~\ref{eq:fe_limit2}). In particular, we focus on the logarithmic corrections to the free energy 
of the mixture of charged rods as already described using scaling analysis in section ~\ref{sec:scaling_section}. 
The electrostatic component of the free energy is given by
 \begin{eqnarray}
 \frac{F_e\left\{L_k \rightarrow \infty \right\}}{k_BT}  &=& \frac{\Omega}{2}\int \frac{d^{3}\mathbf{q}}{(2\pi)^3} \left\{\ln \left[1 + \frac{\kappa^2 }{q^2} + \frac{\pi \kappa_p^2}{q^3}\right]  - \frac{\kappa^2 }{q^2} - \frac{\pi \kappa_p^2}{q^3} \right \}
 \end{eqnarray}
For weakly charged rods so that $\kappa_p^2 \ll 1$, 
 \begin{eqnarray}
 \frac{F_e\left\{L_k \rightarrow \infty \right\}}{k_BT}  &=& \frac{\Omega}{2}\int \frac{d^{3}\mathbf{q}}{(2\pi)^3} \left\{\ln \left[1 + \frac{\kappa^2 }{q^2} \right]  - \frac{\kappa^2 }{q^2}  \right \}-  \frac{\Omega}{2}\int\frac{d^{3}\mathbf{q}}{(2\pi)^3} \frac{\pi \kappa_p^2 \kappa^2}{q^3 \left(q^2 + \kappa^2\right)} \nonumber \\
 && - \frac{\Omega}{4}\int \frac{d^{3}\mathbf{q}}{(2\pi)^3} \frac{\pi^2 \kappa_p^4}{q^2 \left(q^2 + \kappa^2\right)^2} + \cdots \nonumber \\
 &\simeq& -\frac{\Omega \kappa^3}{12 \pi}
 - \frac{\Omega \kappa_p^2}{4 \pi}\ln \left[\frac{\kappa}{q}\right]_{q\rightarrow 0} - \frac{\Omega \pi \kappa_p^4}{32 \kappa^3} + \cdots,
 \end{eqnarray}
which is Eq. ~\ref{eq:fe_limit1}. On the other hand, for strongly charged rods in weakly screened solutions $\kappa_p^2 \gg 1$ and $\kappa \rightarrow 0$. In this limit, 
 \begin{eqnarray}
 \frac{F_e\left\{L_k \rightarrow \infty \right\}}{k_BT}  &\simeq& \frac{\Omega}{2}\int \frac{d^{3}\mathbf{q}}{(2\pi)^3} \left \{\ln \left[1 + \frac{\kappa^2 }{q^2} \right]  - \frac{\kappa^2 }{q^2}\right\} \nonumber \\
&& +  \frac{\Omega}{2}\int \frac{d^{3}\mathbf{q}}{(2\pi)^3} \left\{\ln \left[1 + \frac{\pi \kappa_p^2}{q^3 }\left\{ 1 - \frac{\kappa^2}{q^2} + \frac{\kappa^4}{q^4}\right\}\right] - \frac{\pi \kappa_p^2}{q^3 } \right \}\nonumber \\
 &=& -\frac{\Omega \kappa^3}{12 \pi} +  \frac{\Omega}{2}\int \frac{d^{3}\mathbf{q}}{(2\pi)^3} \left\{\ln \left[1 + \frac{\pi \kappa_p^2}{q^3 }\right]  - \frac{\pi \kappa_p^2}{q^3 } \right \} + \frac{\Omega}{2}\int \frac{d^{3}\mathbf{q}}{(2\pi)^3} \ln \left[ 1 - \frac{\frac{\kappa^2}{q^2} - \frac{\kappa^4}{q^4}}{1 + \frac{\pi \kappa_p^2}{q^3 }}\right] \nonumber \\
 &=& -\frac{\Omega \kappa^3}{12 \pi} + \frac{\Omega}{2}\int \frac{d^{3}\mathbf{q}}{(2\pi)^3} \left\{\ln \left[1 + \frac{\pi \kappa_p^2}{q^3 }\right] - \frac{\pi \kappa_p^2}{q^3 } \right \}  - \frac{\Omega}{2}\int \frac{d^{3}\mathbf{q}}{(2\pi)^3} \frac{\frac{\kappa^2}{q^2} - \frac{\kappa^4}{q^4}}{1 + \frac{\pi \kappa_p^2}{q^3 }} + \cdots \nonumber \\
 &=& -\frac{\Omega \kappa^3}{12 \pi} - \frac{\Omega \kappa_p^2}{6 \pi} \ln \left[\frac{\sqrt{\pi}\kappa_p}{q}\right]_{q\rightarrow 0}  + \frac{\Omega \kappa^2 \kappa_p^{2/3}}{6\sqrt{3}\pi^{2/3}}  - \frac{\Omega \kappa^4}{6\sqrt{3}\pi^{4/3}\kappa_p^{2/3}}- \frac{\Omega \kappa^2}{4 \pi^2} \left[q\right]_{q\rightarrow \infty} + \cdots ,\nonumber \\
 &&
 \end{eqnarray}
which is Eq. ~\ref{eq:fe_limit2}. 
\renewcommand{\theequation}{C-\arabic{equation}}
  \setcounter{equation}{0}  
 \section*{APPENDIX C : Partition function for mixtures of flexible polyelectrolytes } \label{app:C}
Here, we present the derivation of
the free energy for mixtures of oppositely charged flexible polyelectrolytes in the
presence of their counterions. The flexible polyelectrolytes are represented by continuous 
curves so that $\mathbf{R}_{jk}(s_j)$ is the position vector of the $s^{th}$ monomer 
along the $j^{th}$ chain of type $k$. Furthermore, the contour lengths of chains of type $k$ 
is taken to be $L_k = N_k l_k$, $l_k$ being the Kuhn's segment length. Accounting for the conformational 
degrees of freedom of the flexible chains by the path integral representation, 
the partition function for this system can
be written as
\begin{eqnarray}
    Z  &=&  \frac{\prod_{k=+,-}\prod_{j=1}^{n_k} \int D\left[\mathbf{R}_{jk}\right]\exp\left[-\frac{H_0}{k_B T}\right]}{{\prod_{k=+,-}n_k!\prod_{j' =  c+,c-}n_{j'}!}}\prod_{j' =  c+,c-}\prod_{j'' =  1}^{n_{j'}}\int d\mathbf{r}_{j''} \exp \left [- \frac{H_{int}}{k_BT} \right ].  
\end{eqnarray}
where
\begin{eqnarray}
        \frac{H_{0}}{k_BT} &=&  \sum_{k=+,-}\sum_{j=1}^{n_k}\frac{3}{2l_k^2}\int_{0}^{N_k}ds_j \left(\frac{\partial \mathbf{R}_{jk}(s_j)}{\partial s_j}\right)^2 \\
       \frac{H_{int}}{k_BT} &=&  \frac{H_{w}}{k_BT} + \frac{H_{e}}{k_BT} \\
      \frac{H_{w}}{k_BT} &=& \frac{w}{2}\int d\mathbf{r} \left[\hat{\rho}_+(\mathbf{r}) + \hat{\rho}_-(\mathbf{r})\right]^2 \\
\frac{H_{e}}{k_BT} &=& \frac{l_B}{2}\int d\mathbf{r} \int d\mathbf{r}' \frac{\hat{\rho}_e(\mathbf{r})\hat{\rho}_e(\mathbf{r}')}{|\mathbf{r} - \mathbf{r}'|}.
\end{eqnarray}
Note here that in contrast to the rodlike polymers, the Hamiltonian for the flexible polyelectrolytes 
has an additional contribution coming from the chain connectivity ($H_0$) in 
addition to the contributions coming from short range excluded volume ($H_w$) and long range electrostatic 
interactions ($H_e$). Furthermore, in the spirit of polymer field theories\cite{glenn_book}, we have 
replaced $H_w$ by the delta functional form and defined a excluded volume parameter $w$ to take care of the 
short range interactions. Also, the microscopic densities in the above equations are defined as
        \begin{eqnarray}
      \hat{\rho}_e(\mathbf{r}) &=& Z_{+}\sigma_+ \hat{\rho}_+(\mathbf{r}) + Z_{-}\sigma_- \hat{\rho}_-(\mathbf{r}) + \sum_{j'=c+,c-} Z_{j'} \hat{\rho}_{j'}(\mathbf{r}) \\
    \hat{\rho}_{k}(\mathbf{r})  &=& \sum_{j=1}^{n_{k}}\int_{0}^{N_k} ds_{j} \, \delta \left[\mathbf{r} - \mathbf{R}_{jk}(s_j)\right], \quad \mbox{for} \quad k = +,-,\\
     \hat{\rho}_{j'}(\mathbf{r}) &=&  \sum_{j''=1}^{n_{j'}} \delta (\mathbf{r}-\mathbf{r}_{j''}), \quad \mbox{for} \quad j' = c+,c-.
\end{eqnarray}
To proceed further, we rewrite the partition function as 
\begin{eqnarray}
       Z& = & Z_o \left< \exp \left [- \frac{H_{int}}{k_BT} \right ]\right>_{R_+,R_-} \label{eq:parti_flex}
\end{eqnarray}
where $Z_o$ is the partition function of a mixture of non-interacting chains and counterions. Explicitly, it is 
given by 
\begin{eqnarray}
      Z_o &=& \frac{\prod_{k=+,-}\prod_{j=1}^{n_k}  \int D\left[\mathbf{R}_{jk}\right]\exp\left[-\frac{H_0}{k_BT}\right]}{\prod_{k=+,-}n_k!\prod_{j'=c+,c-}n_{j'}!} \prod_{j' =  c+,c-}\prod_{j'' =1}^{n_{j'}}\int d\mathbf{r}_{j''}.
      \end{eqnarray}
Using the Stirling's approximation $n! \simeq n \ln n - n$, we get 
 \begin{eqnarray}
\frac{F_o}{k_BT} & = &  - \ln Z_o  = -\sum_{k=+,-} n_k \ln Q_k^0 + \sum_{\gamma =  +,-,c+,c-} n_{\gamma} \left[\ln \frac{n_{\gamma}}{\Omega} - 1\right],
\end{eqnarray}
where $F_o$ is the Helmholtz free energy of the mixtures of non-interacting chains and counterions. 
Also, $Q_k^0$ is the partition function of a \emph{single} Gaussian chain of type $k$, given by 
\begin{eqnarray}
      Q_k^0 &=& \frac{1}{\Omega}\int D\left[\mathbf{R}_{k}\right]\exp\left[-\frac{3}{2l_k^2}\int_{0}^{N_k}ds \left(\frac{\partial \mathbf{R}_{k}(s)}{\partial s}\right)^2\right].
\end{eqnarray}
In writing the above equation, we have dropped the index representing the chain number in the path 
integral. Furthermore,
\begin{eqnarray}
    \left<\exp \left [- \frac{H_{int}}{k_BT} \right ]\right>_{R_+,R_-} &=& \frac{ \prod_{k}\prod_{j=1}^{n_k} \int D\left[\mathbf{R}_{jk}\right]\exp\left[-\frac{H_0}{k_BT}\right]\prod_{j'}\prod_{j''=1}^{n_{j'}}\int d\mathbf{r}_{j''} \exp \left [- \frac{H_{int}}{k_BT} \right ] }{ \prod_{k}\prod_{j=1}^{n_k}\int D\left[\mathbf{R}_{jk}\right]\exp\left[-\frac{H_0}{k_BT}\right]\prod_{j'}\prod_{j''=1}^{n_{j'}}\int d\mathbf{r}_{j''}}, \nonumber \\
    &&
\end{eqnarray}
where $k=+,-$ and $j' =  c+,c-$. Using Hubbard-Stratonovich transformation\cite{glenn_book} for the
excluded volume and electrostatic parts in the Hamiltonian, and using three dimensional 
Fourier transforms as defined in Appendix A, the partition function can be written as 
\begin{eqnarray}
       \frac{Z}{Z_o'} &=&  \frac{ \int \prod_{q}\left[D[\phi_q] D[\psi_q]\right]\exp\left[-\frac{1}{2}\int \frac{d^{3}\mathbf{q}}{(2\pi)^3}\left[\phi_q \frac{1}{w}\phi_{-q} + \psi_q \frac{q^2}{4\pi l_B}\psi_{-q} \right ]\right]}{\int \prod_{q}\left[D[\phi_q] D[\psi_q]\right]\exp\left[-\frac{1}{2}\int \frac{d^{3}\mathbf{q}}{(2\pi)^3}\left(\phi_q \frac{1}{w}\phi_{-q} + \psi_q \frac{q^2}{4\pi l_B}\psi_{-q}\right)\right ] }\times \nonumber \\
       && \exp\left[ \sum_{k=+,-}n_k \ln Q_k\left\{\phi_q,\psi_q \right\} + \sum_{j'=c+,c-}n_{j'}\ln Q_{j'}\left\{\psi_q\right\}\right]
\end{eqnarray}
where
\begin{eqnarray}
       Z_o' &=&  Z_o\exp\left[-\frac{w}{2 \Omega}\left( n_+ N_+ + n_- N_-\right)^2 \right ]
\end{eqnarray}
and  $Q_{j'}$ is the partition function for a single small ion of type $j'$, given by Eq. ~\ref{eq:partition_smallions}. 
Also, $Q_k$ is the partition function for a single chain of type $k$, given by
\begin{eqnarray}
       Q_{k}\left\{\phi_q,\psi_q \right\} & = & \frac{\int D\left[\mathbf{R}_k\right]\exp\left[-\frac{3}{2l_k^2}\int_{0}^{N_k}ds \left(\frac{\partial \mathbf{R}_k(s)}{\partial s}\right)^2 + i \int \frac{d^{3}\mathbf{q}}{(2\pi)^3}\left( \phi_q + Z_k \sigma_k \psi_q \right) \rho_{k,-q} \right]}{\int D\left[\mathbf{R}_k\right]\exp\left[-\frac{3}{2l_k^2}\int_{0}^{N_k}ds \left(\frac{\partial \mathbf{R}_k(s)}{\partial s}\right)^2\right]}.\nonumber \\
       &&
\end{eqnarray}
Here, $\rho_{k,q}$ is the Fourier component of the microscopic density of a \textit{single} chain 
of type $k$, given by
  \begin{eqnarray}
    \rho_{k,q}  &=& \int_{0}^{N_k} ds e^{i\mathbf{q}\cdot\mathbf{R}_k(s)}, \quad \mbox{for} \quad k = +,-.
\end{eqnarray}
Expanding in powers of $\rho_{k,q}$ and using translational invariance
\begin{eqnarray}
       Q_{k} & \simeq & 1 - \frac{1}{2\Omega} \int \frac{d^{3}\mathbf{q}}{(2\pi)^3}\left( \phi_q + Z_k \sigma_k \psi_q \right) \left<\rho_{k,q} \rho_{k,-q}\right> \left( \phi_{-q} + Z_k \sigma_k \psi_{-q} \right) \nonumber \\
&=& 1 - \frac{1}{2\Omega} \int \frac{d^{3}\mathbf{q}}{(2\pi)^3}\left( \phi_q + Z_k \sigma_k \psi_q \right) N_k^2g\left(q^2N_kl_k^2/6 \right)\left( \phi_{-q} + Z_k \sigma_k \psi_{-q} \right) \nonumber \\
& \simeq & \exp \left[- \frac{1}{2\Omega} \int \frac{d^{3}\mathbf{q}}{(2\pi)^3}\left( \phi_q + Z_k \sigma_k \psi_q \right) N_k^2g\left(q^2N_kl_k^2/6 \right)\left( \phi_{-q} + Z_k \sigma_k \psi_{-q} \right) \right]
\end{eqnarray}
where $g(x)$ is Debye function, given by
\begin{eqnarray}
       g(x) &=& \frac{2}{x^2}\left[e^{-x} - 1 + x\right]
\end{eqnarray}
Consider the special case of \textbf{symmetric} mixtures so that $Z_+ \sigma_+ n_+ N_+^2 g\left(q^2N_+ l_+^2/6 \right) = -Z_- \sigma_- n_- N_-^2 g\left(q^2N_-l_-^2/6 \right)$. For this particular case, the cross terms containing $\phi \psi$ vanishes. Using the approximation for the partition function of small ions as given by Eq. ~\ref{eq:log_exp}, 
the partition function for the symmetric mixtures of oppositely charged flexible polyelectrolytes becomes
\begin{eqnarray}
       \frac{Z}{Z_o'} &=&  \frac{ \int \prod_{q}D[\phi_q] \exp\left[-\frac{1}{2}\int \frac{d^{3}\mathbf{q}}{(2\pi)^3}\phi_q \left\{\frac{1}{w} + \frac{1}{\Omega}\sum_{k=+,-}N_k^2 n_k g\left(q^2N_k l_k^2/6\right)\right\}\phi_{-q}\right ] }{\int \prod_{q}D[\phi_q]\exp\left[-\frac{1}{2}\int \frac{d^{3}\mathbf{q}}{(2\pi)^3}\phi_q \frac{1}{w} \phi_{-q}\right ] }\times \nonumber \\
&&  \frac{ \int \prod_{q}D[\psi_q] \exp\left[-\frac{1}{8\pi l_B}\int \frac{d^{3}\mathbf{q}}{(2\pi)^3}\psi_q \left\{q^2 + \kappa^2
+ \sum_{k=+,-}\kappa_k^2 g\left(q^2N_k l_k^2/6\right)\right\}\psi_{-q}\right ] }{\int \prod_{q}D[\psi_q]\exp\left[-\frac{1}{8\pi l_B}\int \frac{d^{3}\mathbf{q}}{(2\pi)^3}\psi_q q^2 \psi_{-q}\right ] } \nonumber \\
&&
\end{eqnarray}
where $\kappa^2 = 4\pi l_B \sum_{j'=c+,c-} Z_{j'}^2 n_{j'}/\Omega $ and $\kappa_k^2 = 4\pi l_B Z_{k}^2 \sigma_{k}^2 N_k^2 n_k/\Omega$ are the same as defined in Appendix A for rodlike polyelectrolytes. Carrying out the Gaussian functional integrals and subtracting out the free energy in the low density limit, Eq. ~\ref{eq:free_flex} is obtained.

\section*{REFERENCES}
\setcounter {equation} {0}
\pagestyle{empty} \label{REFERENCES}

\newpage

\vspace*{1.0cm}

\vspace*{1.0cm}

\newpage

\begin{figure}[ht!]
\vspace*{3.0cm}
\begin{center}
    \includegraphics[height=3.6cm,width=4.6cm]{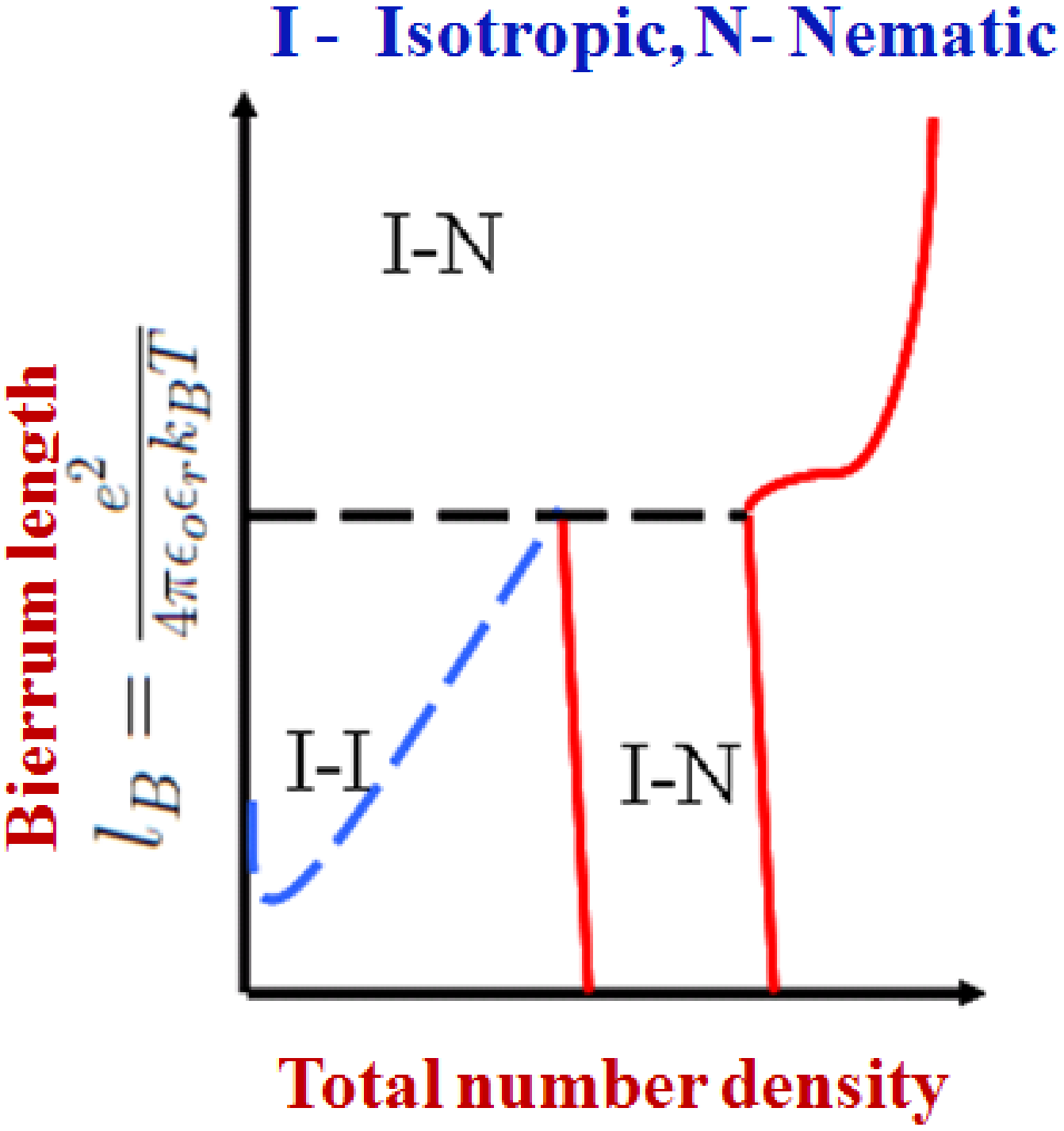}
\caption{For table of contents use only}
\end{center}
\end{figure}


\begin{thebibliography}{99}

\bibitem{dubin_review}
Cooper, C. L.; Dubin, P. L.; Kayitmazer, A. B.; Turksen, S. Current Opinion in Colloid and Interface Science 2005, 10, 52.

\bibitem{pe_complex_reviews}
Thunemann, A. F.; Muller, M.; Dautzenberg, H.; Joanny, J. F. O.; Lowne, H. Polyelectrolyte complexes. In Polyelectrolytes with Defined Molecular Architecture Ii, 2004; Vol. 166; pp 113.

\bibitem{vlad_muthu}
 Belyi, V. A.; Muthukumar, M. Proceedings of the National Academy of Sciences of the United States of America 2006, 103, 17174.

\bibitem{shklovskii_virus}
Hu, T.; Shklovskii, B. I. Physical Review E 2007, 75, 051901.

\bibitem{waite}
Waite, J. H.; Andersen, N. H.; Jewhurst, S.; Sun, C. J. Journal of Adhesion 2005, 81, 297.

\bibitem{bazan_biosensor}
Hong, J. W.; Henme, W. L.; Keller, G. E.; Rinke, M. T.; Bazan, G. C. Advanced Materials 2006, 18, 878.

\bibitem{voorn_review}
Voorn M.J. Advances in Polymer Science - Fortschritte Der Hochpolymeren-Forschung  (Springer Berlin / Heidelberg, 1959 ),
(1),192-233.

\bibitem{veis_complexation}
Veis, A. in \textit{Biological Polyelectrolytes}(Marcel Dekker Inc., New York, 1970).

\bibitem{kabanov_work1}
Bakeev,K. N.; Izumrudov, V. A.; Kuchanov, S. I.; Zezin, A. B.; Kabanov, V. A. Macromolecules
1992, 25, 4249.

\bibitem{kabanov_work2}
Kabanov, A. V.; Bronich, T. K.; Kabanov, V. A.; Yu, K.; Eisenberg,A. Macromolecules 1996, 29, 6797. 

\bibitem{dautzenberg_work1}
Dautzenberg, H.; Hartmann, J.; Grunewald, S.; Brand, F. Ber. Bunsen-Ges. Phys. Chem. 1996, 100, 1024. 

\bibitem{pogodina_work}
Pogodina, N. V.; Tsvetkov,N. V. Macromolecules 1997, 30, 4897. 

\bibitem{dautzenberg_work2}
Dautzenberg, H. Macromolecules 1997, 30,7810.

\bibitem{zhaoyang_complex}
Ou, Z. Y.; Muthukumar, M. Journal of Chemical Physics 2006, 124, 11.


\bibitem{voorn_overbeek}
 Michaeli, I.; Overbeek, J. T. G.; Voorn, M. J. Journal of Polymer Science 1957, 23, 443.

\bibitem{voorn_1}
Voorn, M. J. Recueil Des Travaux Chimiques Des Pays-Bas-Journal of the Royal Netherlands Chemical Society 1956, 75, 317.

\bibitem{voorn_2}
Voorn, M. J. Recueil Des Travaux Chimiques Des Pays-Bas-Journal of the Royal Netherlands Chemical Society 1956, 75, 405.

\bibitem{voorn_3}
Voorn, M. J. Recueil Des Travaux Chimiques Des Pays-Bas-Journal of the Royal Netherlands Chemical Society 1956, 75, 427.

\bibitem{voorn_4}
Voorn, M. J. Recueil Des Travaux Chimiques Des Pays-Bas-Journal of the Royal Netherlands Chemical Society 1956, 75, 925.

\bibitem{voorn_5}
Voorn, M. J. Recueil Des Travaux Chimiques Des Pays-Bas-Journal of the Royal Netherlands Chemical Society 1956, 75, 1021.

\bibitem{borue_complexation}
Borue, V. Y.; Erukhimovich, I. Y. Macromolecules 1990, 23, 3625.

\bibitem{monica_complexation}
Kudlay, A.; de la Cruz, M. O. Journal of Chemical Physics 2004, 120, 404.

\bibitem{monica_complexation_2}
Kudlay, A.; Ermoshkin, A. V.; de la Cruz, M. O. Macromolecules 2004, 37, 9231.

\bibitem{joanny_01}
Castelnovo, M.; Joanny, J. F. European Physical Journal E 2001, 6, 377.

\bibitem{jay_yuri_1}
Popov, Y. O.; Lee, J. H.; Fredrickson, G. H. Journal of Polymer Science Part B-Polymer Physics 2007, 45, 3223.

\bibitem{jay_yuri_2}
Lee, J.; Popov, Y. O.; Fredrickson, G. H. Journal of Chemical Physics 2008, 128, 224908.

\bibitem{muthu_double}
Muthukumar, M. Journal of Chemical Physics 1996, 105, 5183.

\bibitem{degennes_prost}
de Gennes, P.G.; Prost J. \textit{The Physics of Liquid Crystals} (Clarendon Press, Oxford, 1993).

\bibitem{degennes_pincus}
de Gennes, P. G.; Pincus, P.; Velasco, R. M.; Brochard, F. Journal De Physique 1976, 37, 1461.

\bibitem{joanny_leibler}
Joanny, J. F.; Leibler, L. Journal De Physique 1990, 51, 545.

\bibitem{khokhlov_nyrkova}
Khokhlov, A. R.; Nyrkova, I. A. Macromolecules 1992, 25, 1493.

\bibitem{muthu_weak_polymer}
Muthukumar, M. Macromolecules 2002, 35, 9142.

\bibitem{chilun_muthu}
Lee, C. L.; Muthukumar, M. Journal of Chemical Physics 2009, 130, 024904.

\bibitem{onsager_paper}
Onsager, L. Annals of the New York Academy of Sciences 1949, 51, 627.

\bibitem{odijk_pe_rods}
Stroobants, A.; Lekkerkerker, H. N. W.; Odijk, T. Macromolecules 1986, 19, 2232.

\bibitem{chen_koch}
Chen, S. B.; Koch, D. L. Journal of Chemical Physics 1996, 104, 359.

\bibitem{carri_muthu99}
Carri, G. A.; Muthukumar, M. Journal of Chemical Physics 1999, 111, 1765.

\bibitem{potemkin_1}
Potemkin, II; Limberger, R. E.; Kudlay, A. N.; Khokhlov, A. R. Physical Review E 2002, 66, 011802.

\bibitem{potemkin_2}
Potemkin, II; Khokhlov, A. R. Journal of Chemical Physics 2004, 120, 10848.

\bibitem{potemkin_3}
Potemkin, II; Oskolkov, N. N.; Khokhlov, A. R.; Reineker, P. Physical Review E 2005, 72, 021804.

\bibitem{stell_rods}
Hoye, J.; Raineri, F. O.; Stell, G.; Routh, J. Journal of Statistical Physics 2003, 110, 835.

\bibitem{parsegian}
Brenner, S. L.; Parsegia.V.A. Biophysical Journal 1974, 14, 327.

\bibitem{edwardsbook}
Doi, M.; Edwards, S.F. \textit{The Theory of Polymer Dynamics} (Clarendon Press, Oxford, 1986), chapter 10.

\bibitem{flory_semiflex}
Flory, P. J. Proceedings of the Royal Society of London Series a-Mathematical and Physical Sciences 1956, 234, 73.

\bibitem{flory_rods}
Flory, P. J. Proceedings of the Royal Society of London Series a-Mathematical and Physical Sciences 1956, 234, 60.

\bibitem{khokhlov_semiflex}
Khokhlov, A. R. Physics Letters A 1978, 68, 135.

\bibitem{khokhlov_semenov}
Khokhlov, A. R.; Semenov, A. N. Physica A 1981, 108, 546.

\bibitem{rubinstein_review}
Dobrynin, A. V.; Rubinstein, M. Progress in Polymer Science 2005, 30, 1049.

\bibitem{kaji_kanaya}
Nishida, K.; Kaji, K.; Kanaya, T. Journal of Chemical Physics 2001, 114, 8671.

\bibitem{odijk_self}
Odijk, T. Journal of Physical Chemistry 1989, 93, 3888.

\bibitem{safran_pincus}
Mackintosh, F. C.; Safran, S. A.; Pincus, P. A. Europhysics Letters 1990, 12, 697.

\bibitem{onsager_trial_func}
Franco-Melgar, M.; Haslam, A. J.; Jackson, G. Molecular Physics 2008, 106, 649.

\bibitem{manning}
Manning, G. S. Journal of Chemical Physics 1969, 51, 924.

\bibitem{lakatos_legendre}
Lakatos K. Journal of Statistical Physics 1970,2, 121.

\bibitem{lasher_legendre}
Lasher, G. Journal of Chemical Physics 1970, 53, 4141.

\bibitem{intgral_equation_real}
Herzfeld, J.; Berger, A. E.; Wingate, J. W. Macromolecules 1984, 17, 1718.

\bibitem{numerical_recipe}
Press, W.H. et al. \textit{Numerical Recipes in C}(Cambridge University Press,New York, 1992).

\bibitem{bifurcation}
Kayser, R. F.; Raveche, H. J. Physical Review A 1978, 17, 2067.

\bibitem{chandler_book}
Chandler, D. \textit{Introduction to Modern Statistical Mechanics}(Oxford University Press, New York, 1987).


\bibitem{fisher_levin}
Fisher, M. E.; Levin, Y. Physical Review Letters 1993, 71, 3826.

\bibitem{fisher}
Fisher, M. E. Journal of Statistical Physics 1994, 75, 1.

\bibitem{glenn_book}
Fredrickson, G.H. \textit{The Equilibrium Theory of Inhomogeneous Polymers} (Oxford University Press, New York, 2006).

\bibitem{leibler_borukhov}
Borukhov, I.; Andelman, D.; Borrega, R.; Cloitre M.; Leibler, L.; Orland, H. Journal of Physical Chemistry B 2000, 104 11027.

\bibitem{straley_review}
 Stephen,M.J.; Straley, J.P. Reviews of Modern Physics 1974, 46, 617. 

\bibitem{parsons_paper}
Parsons, J.D. Physical Review A 1979, 19, 1225.

\bibitem{lee_higher}
Lee, S.D. Journal of Chemical Physics 1987, 87, 4972.

\bibitem{levin_charge_regularization}
Diehl, A.; Barbosa, M.C.; Levin, Y. Europhysics Letters 2001, 53, 86.   

\bibitem{muthu_charge_regularization}
Muthukumar, M.; Hua, J.; Kundagrami, A. Journal of Chemical Physics 2010, 132, 084901.

\bibitem{rubinstein_pairing}
Wang, Z.W.; Rubinstein, M. Macromolecules 2006, 39, 5897.                  

\bibitem{ebeling_work}
Falkenhagen, H.; Ebeling, W. in \textit{Ionic Interactions}, edited by S. Petrucci (Academic, New York, 1971), vol. 1.

\bibitem{levin_review}
Levin, Y. Reports on Progress in Physics 2002, 65, 1577.

\bibitem{counterion_adsorption_muthu}
Muthukumar, M. Journal of Chemical Physics 2004, 120, 9343; Kumar, R.; Kundagrami, A.; Muthukumar, M. Macromolecules 2009, 42, 1370.    

\bibitem{semenov_rubinstein}
Semenov, A.N.; Rubinstein, M. Macromolecules 1998, 31, 1373.

 


\end{thebibliography}
\end{document}